\documentclass[journal]{IEEEtran}
\usepackage[utf8]{inputenc}
\usepackage{amsmath}
\usepackage{amsthm}
\usepackage{multirow}
\usepackage{amssymb}
\usepackage{graphicx}
\usepackage{float}
\usepackage{bm}

\usepackage{cite}
\usepackage{comment}
\usepackage{epstopdf}
\usepackage{float}

\newenvironment{Procedure}[1][htb]
  {
   \begin{algorithm}[#1]%
  }{\end{algorithm}}

\DeclareMathOperator*{\argmax}{arg\,max}
\usepackage{amsfonts}
\usepackage{amsthm}
\usepackage{subcaption}
\usepackage{tikz}

\usepackage[noend]{algpseudocode}
\usepackage[ruled,norelsize,linesnumbered]{algorithm2e}
\usepackage{multicol}
\usepackage{circledsteps}

\algdef{SE}[DOWHILE]{Do}{doWhile}{\algorithmicdo}[1]{\algorithmicwhile\ #1}%
\makeatletter
\def\BState{\State\hskip-\ALG@thistlm}

\newlength\myindent
\newtheorem{remark}{Remark}
\makeatletter
\newcommand{\removelatexerror}{\let\@latex@error\@gobble}
\makeatother

\title{\Large\bf Cloud-Assisted Collaborative Road Information Discovery with Gaussian Process: Application to Road Profile Estimation}
\author{ Mohammad R.~Hajidavalloo,
Zhaojian~Li$^*$,
Xin~Xia,
Ali~Louati,
Minghui~Zheng,
and Weichao Zhuang
\thanks{\quad This work is supported by National Science Foundation grant \#2030411/2030375 and grant \#2045436.}
\thanks{\quad  Mohammad Hajidavalloo and Zhaojian Li are with the Department of Mechanical Engineering, Michigan State University, East Lansing, MI 48824, USA.
        Email: {\tt\small \{hajidava,lizhaoj1\}@egr.msu.edu}.}
\thanks{\quad  Xin Xia is with the Department of Civil and Environmental Engineering, University of California at Los Angeles, CA 90095.
        Email: {\tt\small x35xia@g.ucla.edu}.}
\thanks{\quad  Ali Louati is with Department of Information Systems, College of Computer Engineering and Sciences, Prince Sattam bin Abdulaziz University, Al-Kharj 11942, Saudi Arabia.
        Email: {\tt\small a.louati@psau.edu.sa}.}
\thanks{\quad  Minghui Zheng is with the Department of Mechanical and Aerospace Engineering, University at Buffalo, Buffalo, NY 14260, USA.
        Email: {\tt\small mhzheng@buffalo.edu}.}
\thanks{\quad  Weichao Zhuang is with the School of Mechanical Engineering, Southeast University, Nanjing, Jiangsu, 211189 China. 
        Email: {\tt\small wezhuang@seu.edu.cn}.
       } 
\thanks{$*$ Zhaojian Li is the corresponding author.}
        }%

\IEEEoverridecommandlockouts
\begin{document}
\maketitle
\begin{abstract}
There is an increasing popularity in exploiting modern vehicles as mobile sensors to obtain important road information such as potholes, black ice and road profile. Availability of such information has been identified as a key enabler for next-generation vehicles with enhanced safety, efficiency, and comfort.  However,  existing road information discovery approaches have been predominately performed in a single-vehicle setting, which is inevitably susceptible to vehicle model uncertainty and measurement errors. To overcome these limitations, this paper presents a novel cloud-assisted collaborative  estimation framework that can utilize multiple heterogeneous vehicles to iteratively enhance estimation performance. Specifically,  each vehicle combines its onboard measurements with a cloud-based Gaussian process (GP), crowdsourced from prior participating vehicles as ``pseudo-measurements'',  into a local estimator to refine the estimation. 
The resultant local onboard estimation is then sent back to the cloud to update the GP, where  we utilize a noisy input GP (NIGP) method to explicitly handle uncertain GPS measurements. We employ the proposed framework to the application of collaborative road profile estimation. Promising results on extensive simulations and hardware-in-the-loop experiments  show that the proposed collaborative estimation  can significantly enhance estimation and iteratively improve the performance from vehicle to vehicle, despite vehicle heterogeneity, model uncertainty, and measurement noises.
\end{abstract}
\section{Introduction}
There is a growing interest in employing road information in intelligent vehicle systems to improve road safety \cite{safety}, ride comfort \cite{comfort}, and fuel efficiency \cite{fuel_economy1,fuel_economy2}. Real-time and crowd-sourced road information (e.g., black ice, potholes, and road roughness) can increase situational awareness, enhance control performance, and provide additional functionalities \cite{future_mobility}. Furthermore, road surface monitoring is synergistic with many nations' urgent need to rebuild and modernize the road infrastructure \cite{budget}, by offering the government/agency up-to-date road condition information to best plan  road maintenance. Meanwhile, modern  automotive  vehicles  are  equipped  with  advanced sensing  and  connectivity  capabilities,  which  can be exploited to discover real-time road information 
that is not otherwise easily or economically measurable from dedicated sensors. Such vehicle-based information inference employs system dynamics and can enable efficient and robust road information discovery with a wide road coverage.
 

As such, various vehicle-based estimation approaches have been pursued to exploit vehicle onboard measurements along with the underlying dynamics to reconstruct the road information.
Take road profile estimation as an example,  \textcolor{black}{there are two major classes of road profile estimation schemes:} unknown input observer (UIO)-based \cite{rath2014estimation,li2016simultaneous,li2019optimization} and  extended state observer (ESO)-based \cite{qin2017road,wang2017road,fauriat2016estimation}. The ESO methods exploit an augmented state by treating the road signal as an additional state, which is estimated along with the original states using the commonly used state observers such as Kalman Filter (KF) for linear systems and high gain observers (HGO) or extended KF (EKF) for nonlinear dynamic systems \cite{li2019optimization,qin2017road,wang2017road,fauriat2016estimation,zheng2016extended,zheng2016multi}.
\textcolor{black}{In this regard, the authors in \cite{doumiati2011estimation} proposed an embedded observer based on Kalman filter with two stages. In the first stage the vertical acceleration signal is utilized  to determine the vehicle body position, while in the second stage, a Kalman filter uses the first stage's output as a measure to estimate the road profile elevation and subsequently evaluate loads on the wheels. In \cite{zhao2019road,xue2020road} a road profile estimation method is constructed by utilizing only a smartphone to measure the responses of an ordinary vehicle. Genetic algorithm is first used to identify the model of a half car model, and with the estimated vehicle model, an augmented Kalman filter is designed to estimate the road profile. Rauch-Tung-Streiber (RTS)
smoothing is employed to improve the accuracy.}

\textcolor{black}{On the other hand, the UIO methods generally aim to obtain a precise and stable model inverse to estimate the road information (i.e., the input) from outputs of the system. 
In this regard, in \cite{doumiati2017road}, the road profile is characterized as a sinusoidal disturbance signal acting on the vehicle
system. To handle the unknown and time-varying characteristics of the signal, the presented road profile estimate approach used the internal model concept and the Youla–Kuera (YK) parameterization methodology to construct an adaptive control system (also known as Q-parameterization). In \cite{liu2020line}, the authors used an inverse model to estimate the unknown input (i.e. the road profile signal) by measuring the unsprung mass acceleration, however no adjustments for the case where the measurement is noisy is not provided.
In \cite{rath2014simultaneous}, an integrated nonlinear model is created by combining the vertical and longitudinal dynamics of a quarter wheel. The time-varying random road profile and tire friction are handled as unknown inputs in the modeled dynamics. A combination of nonlinear Lipschitz observer and modified super-twisting algorithm (STA) observer is designed to estimate these unknown inputs and states at the same time.}

\textcolor{black}{
Some other noteworthy works beyond the above two categories are briefly reviewed as follows. Using an artificial neural network (ANN), \cite{yousefzadeh2010road} presents a solution to the problem of estimating road profiles in which the ANN is trained using data from a verified vehicle model in the ADAMS program to estimate road profiles based on accelerations detected by the vehicle.
In \cite{imine2006road}, a novel sliding mode observers-based approach for road profile estimation has been developed and compared to two inertial methods.
In \cite{gohrle2014road}, new methods for generating a preview road height profile using car sensors and controlling the active suspension using this information are developed. Sensor-fixed and inertial coordinate systems are exploited to reconstruct the road height profile from sensor data after establishing a desired road height profile as input for the defined control algorithms.
}

Despite the progresses, the aforementioned approaches are based on a single-vehicle setting and thus susceptible to  model uncertainty and measurement errors. Therefore, the goal of this paper is to develop a new cloud-based collaborative  estimation framework that can exploit multiple heterogeneous vehicles to enhance the accuracy and robustness of road information discovery.
Similar idea of employing cloud to perform consensus-based parameter identification for vehicle diagnostics and prognostics is reported in  \cite{breschi2018cloud}. In addition, a dynamic average consensus-based  distributed cooperative road friction coefficient estimation scheme is presented in \cite{jalalmaab2018cooperative}, which requires real-time vehicle-to-vehicle communication with known and fixed topology that is in general difficult to achieve in practice. In \cite{HuanGao}, an iterative learning-based collaborative road profile estimation is developed. However, it requires a much simplified model to ensure the convergence. 

In this paper, our approach utilizes the cloud as a central platform to crowdsource local vehicle estimations using Gaussian processes (GP \cite{ko2009gp}). The crowdsourced GP is then sent back to the vehicle as an additional ``pseudo-measurement'' to enhance onboard estimation using a local estimator such as Kalman filter. Similar idea of using pseudo-measurements to enhance estimation is also used in state estimation of  power distribution systems, which are high-dimensional with limited available measurements and are thus subject to observability issues; pseudo-measurements are employed therein to improve observability and enhance estimation \cite{jin2021new,dehghan}. The enhanced local estimate is then uploaded to the cloud to update the GP. This process then repeats for the next participating vehicle to iteratively refine the road information estimates.   Note that the cloud-based pseudo-measurements can be pre-loaded to vehicles and the vehicle-based estimates can be sent to cloud only when there is robust network available, without requiring real-time communications and thus making this framework practically appealing.

Furthermore, as everyday vehicles are equipped with GPS of limited accuracy \cite{GPS}, which will inevitably result in noisy correspondence between vehicle position and estimated road information,   we exploit an extended variant of GP,  noisy input Gaussian process (NIGP \cite{mchutchon2011gaussian}),  to explicitly handle the GPS uncertainties.  This is critical to provide road information estimates with satisfactory spatial resolution for practical use. Moreover, we demonstrate this framework in the application of collaborative estimation of road profile,  which has been frequently proposed to be incorporated as a preview to enhance suspension controls for improved safety and comfort \cite{Preview1,Preview2,Preview3}. We show in comprehensive simulations and hardware-in-the-loop experiments that our collaborative estimation framework can greatly enhance robustness and accuracy in the presence of model uncertainty, measurement noise, and vehicle heterogeneity.


The contributions of this paper include the following. First, we develop a novel collaborative estimation framework that systematically integrates cloud-based GP with enhanced local estimation with pseudo-measurements from heterogeneous vehicles to iteratively refine the estimates. 
Second, we explicitly consider the GPS uncertainty existent in the location estimate of everyday vehicles, one of the major hurdles in practical use of vehicle-based road information estimate.  In particular, we employ an NIGP approach that offers improved accuracy and reduced variance in the presence of GPS noises. 
Last but not least, extensive simulations and hardware-in-the-loop experiments are performed to show the efficacy of our framework in the application of collaborative road profile estimation. 

The rest of the paper is organized as follows. In Section II, the problem of collaborative road information discovery is introduced and preliminaries on KF and GP are presented. In Section III, the cloud-assisted collaborative road information discovery framework is detailed and its application to road profile estimation is presented with extensive simulation and experimental results in section IV. Finally, Section~V concludes the paper. 

\section{Problem Statement and Preliminaries}
\subsection{Problem Statement}
\begin{figure}[!h]
    \centering
    \includegraphics[width=0.8\linewidth]{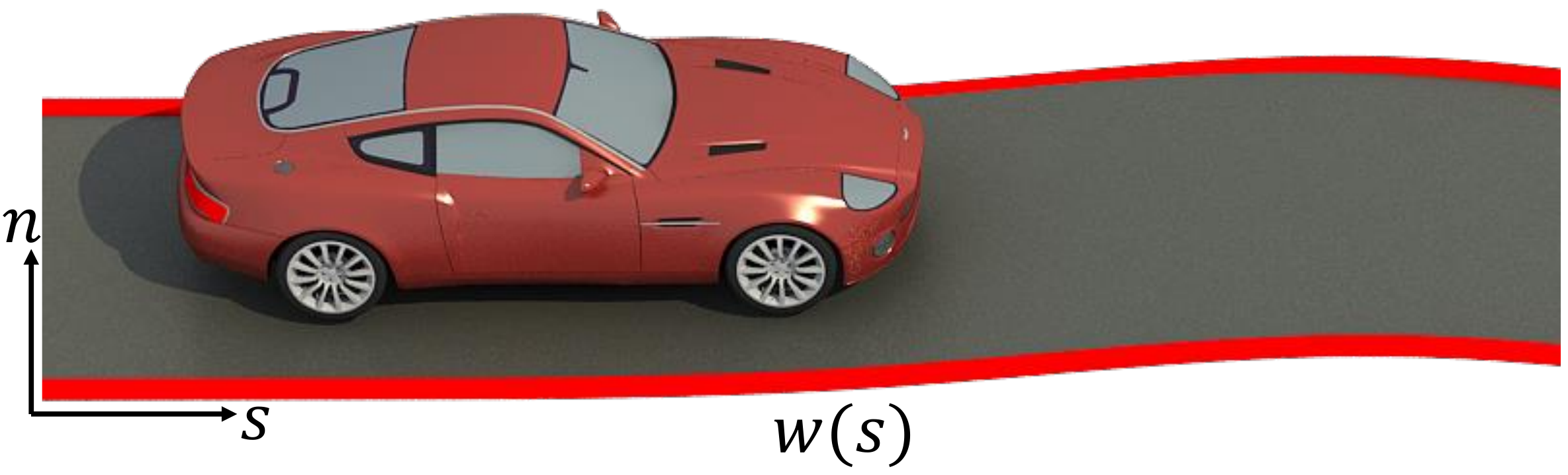}
    \caption{A road segment with road information of interest denoted by $w(s)$.}
    \label{fig:road}
\end{figure}

This paper aims at developing a unified approach to efficiently crowdsource road information from multiple heterogeneous vehicles.
Specifically, given a road segment (e.g., defined by two consecutive   road mile markers \cite{roadmarker}) as illustrated in Fig.~\ref{fig:road}, the  objective of vehicle-based road information estimation  is to use existing onboard sensors (e.g., accelerometers, GPS, yaw rate, roll rate) to discover $w(s)$,  the road information of interest (e.g., road profile, friction coefficient) as a function of distance
in the longitudinal direction (the $s$ direction in 
Fig.~\ref{fig:road}). Here it is assumed that the road condition to be discovered is uniformly distributed in the lateral direction ($n$ direction in Fig.~\ref{fig:road}).  By scaling the distance $s$ with the vehicle speed, the road information to be estimated can also be represented by $w(t) \in \mathbb{R}$, a function of time.

We consider the following discrete-time linear dynamics to characterize the vehicle-road interaction for participating vehicle $i,$ $i=1,2,\cdots$, as:
\begin{equation}
\label{Eq:DT}
\begin{aligned}
&{{\mathbf{x}}}_i(k+1)={A}_i{\mathbf{x}}_i(k)+B_iu_i(k)+D_iw(k)+\eta_i(k), \\
& \mathbf{y}_i(k)={C}_i{\mathbf{x}}_i(k)+{v}_i(k),
\end{aligned}
\end{equation}
where $\mathbf{x}_i\in \mathbb{R}^n$ is the state of the system, $u_i\in\mathbb{R}^m$ is the known control input, $w(k)\in\mathbb{R}$ is the road information of interest (e.g., road profile and road friction coefficient), $\eta_i\in\mathbb{R}^n$ and $v_i\in\mathbb{R}^r$ are system process noise and measurement noise, respectively, $\mathbf{y}_i\in \mathbb{R}^r$ is the measured noisy output, and $A_i$, $B_i$, $C_i$, $D_i$ are system matrices of suitable dimensions.
Similar to \cite{zuo2013energy},  the road information is modeled by a first-order system driven by a white noise:
\begin{equation}
    w(k+1)=aw(k)+be(k),
\end{equation}
where $a$ and $b$ are constant parameters characterizing the road conditions and $e$ is white Gaussian noise. Following the idea of extended state observer \cite{qin2017road,wang2017road,fauriat2016estimation}, one can augment the system state by appending the road information signal $w(k)$ as an additional state, i.e., ${\bar{\mathbf{{x}}}}_i=[\mathbf{x}_{i},w]^T$. As a result, Eqn.~\ref{Eq:DT} can be rewritten as the following augmented discrete-time system:
\begin{equation}
\label{Eq:DT1}
\begin{aligned}
&{\bar{\mathbf{x}}}_i(k+1)=\bar{A}_i\bar{\mathbf{x}}_i(k)+\bar{B}_iu_i(k)+\bar{\eta}_i(k), \\
& \mathbf{y}_i(k)=\bar{C}_i\bar{\mathbf{x}}_i(k)+{v}_i(k).
\end{aligned}
\end{equation}
Here $\bar{A}_i, \bar{B}_i$, and $\bar{C}_i $ are augmented system matrices, and $\bar{\eta}_i=[\eta;e]$ is the augmented process noise. Therefore,   each vehicle can now employ a state estimator (e.g., Kalman filter) to estimate the augmented state which contains the original  state along with the road information of interest.  This is the common practice to estimate the road information parameter based on a single vehicle \cite{li2019optimization,qin2017road,wang2017road,fauriat2016estimation,zheng2016extended,zheng2016multi}. However, such single vehicle-based estimation is inevitably subject to model uncertainty and noisy measurements, which can lead to great errors and large variations. 

As such, the goal of this paper is to develop a cloud-based collaborative estimation framework to iteratively refine the estimates by exploiting multiple heterogeneous vehicles. The idea of this framework is to use the cloud as a central platform to crowdsource estimates from the vehicles using Gaussian process regression (we use an enhanced version to explicitly handle GPS uncertainties) and then use the regressed model as pseudo-measurements to in turn enhance the local estimations (see Section~III for more details). We next briefly review Kalman filter and Gaussian process regression to place our proposed algorithms in later sections in proper context. 

\subsection{Kalman Filter}
The celebrated Kalman filter (KF) is a recursive state estimator for linear systems subject to Gaussian noises, leading to optimal state estimates with minimized mean-square error  \cite{kalman1960new}. 
For the augmented dynamical system represented in (\ref{Eq:DT1}),  at each time step, the KF involves the following two steps:
\begin{equation}
   \begin{aligned}[l]
\label{Eq:KF1}
  & \textbf{Prediction:} \\ 
 & {{{\hat{\bar{\mathbf{x}}}}}_i({k+1|k}})=\bar{{{A}}}_i{{{\hat{\bar{\mathbf{x}}}}}}_i({k|k})+\bar{B}_iu_i(k), \\ 
 & {{{P}}_i({k+1|k}})=\bar{A}_{i}{{{P}}_i({k|k})}\bar{{A}}^{T}_{i}+Q_i. \qquad\qquad\qquad\qquad\qquad\qquad
 \end{aligned}
 \end{equation}
  \begin{equation}
      \begin{aligned}[l]
      \label{Eq:KF2}
      &  \textbf{Correction:} \\
 & {{{\Gamma}}_i({k+1})}={{{P}}_i({k+1|k})}\bar{C}_i^{T}{{[\bar{C}_i{{{P}}_i({k+1|k})}\bar{C}_i^{T}+{R}_i]}^{-1}}, \\ 
 & {{{\hat{\bar{\mathbf{x}}}}}_i({k+1|k+1})}={{{\hat{\bar{\mathbf{x}}}}}_i({k+1|k})}\\
 &\qquad\qquad\qquad+{{{\Gamma}}_i({k+1})}[{{\mathbf{y}}_i({k+1})}-\bar{C}_i{{{\hat{\bar{\mathbf{x}}}}}_i({k+1|k})}], \qquad\qquad\qquad\\ 
 & {{{P}}_i({k+1|k+1})}=[{I}-{{{\Gamma}}_i({k+1})}{{\bar{C}}}_i]{{{P}}_i({k+1|k})}, \\ 
\end{aligned} 
\end{equation}
where ${{{\mathbf{\hat{\bar{x}}}}}_{i}}(k|k):=\mathbb{E}\left\{{{{\mathbf{\bar{x}}}}_{i}(k)}|{{\mathbf{Y}}_{i}}(k)\right\}$ is the state estimate at time step $k$ given all past observations until $k$ with $\mathbf{Y}_{i}(k)=({{\mathbf{y}}_{i}}(1),...,{{\mathbf{y}}_{i}}(k))$,  ${{P}_{i}}(k|k):=\mathbb{E}\{(\mathbf{\bar{x}}_{i}(k)-\mathbf{\hat{\bar{x}}}_{i}(k))(\mathbf{\bar{x}}_{i}(k)-\mathbf{\hat{\bar{x}}}_{i}(k))^{T}|{{\mathbf{Y}}_{i}}(k)\}$ is the covariance matrix at time step $k$, and $\Gamma_i(k+1)$ is referred to as the Kalman gain at time step $k+1$ to correct the estimation based on the measurement-prediction mismatch.
Here $Q_i=\mathbb{E}\{\bar{\eta}_i\bar{\eta}_{i}^{T}\}$ and $R_i=\mathbb{E}\{v_iv_{i}^{T}\}$ denote the process and measurement noise covariances, respectively. With the augmented state estimate, the road information can be recovered as it is a sub-state of $\hat{\bar{\mathbf{x}}}_i$ and
we denote the estimated road information as
$\hat{{w}}_i(k|k)=\mathbb{E}\{w_i(k)|\mathbf{Y}_i(k)\}$.

\subsection{Gaussian Process}\label{sec:GP}
As stated above, the road information of interest can be described by a function of the spatial distance, $w(s)$, or characterized by its power spectrum density \cite{andren2006power}. An alternative description  is from the machine learning perspective using the Gaussian process (GP) model \cite{GP-Intro}, i.e., $w(s)\sim \mathcal{GP}(m(s),\mathcal{K}(s,s'))$, where $m(s)=\mathbb{E}\{w(s)\}$ is the mean function that can take the form of $m(s)=\sum_{j=1}^K\beta_j\psi_j(s):=\beta^\text{T}\psi(s)$. Here $\psi(\cdot)$ is the vector of $K$ basis functions (e.g., polynomial functions or Gaussian basis functions), and $\beta$ is the vector of corresponding linear weights to be trained from data. The kernel function,   $\mathcal{K}(s,s')=cov(s,s')$, 
 characterizes the covariance between any two spatial points $s$ and $s'$, an example of which is the exponentiated quadratic kernel $\mathcal{K}(s,s')=\sigma^2\exp\left(-\frac{\|s-s'\|^2}{2l^2}\right)$, where $\sigma$ and $l$ are the hyper-parameters representing the standard deviation and the lengthscale, respectively. We denote by $\Theta$ the set of hyper-parameters from the mean function and the kernel function  (e.g., $\Theta=\{\beta,\sigma,l\}$ for the above example functions). This GP representation of road information parameter is advantageous as it not only provides an estimated value (i.e., the mean function) but also provides the estimation uncertainties characterized by the kernel function. In this paper, we use GP to crowdsource the estimates from multiple heterogeneous vehicles, which we present next.


\begin{figure}
    \centering
    \includegraphics[width=0.8\linewidth]{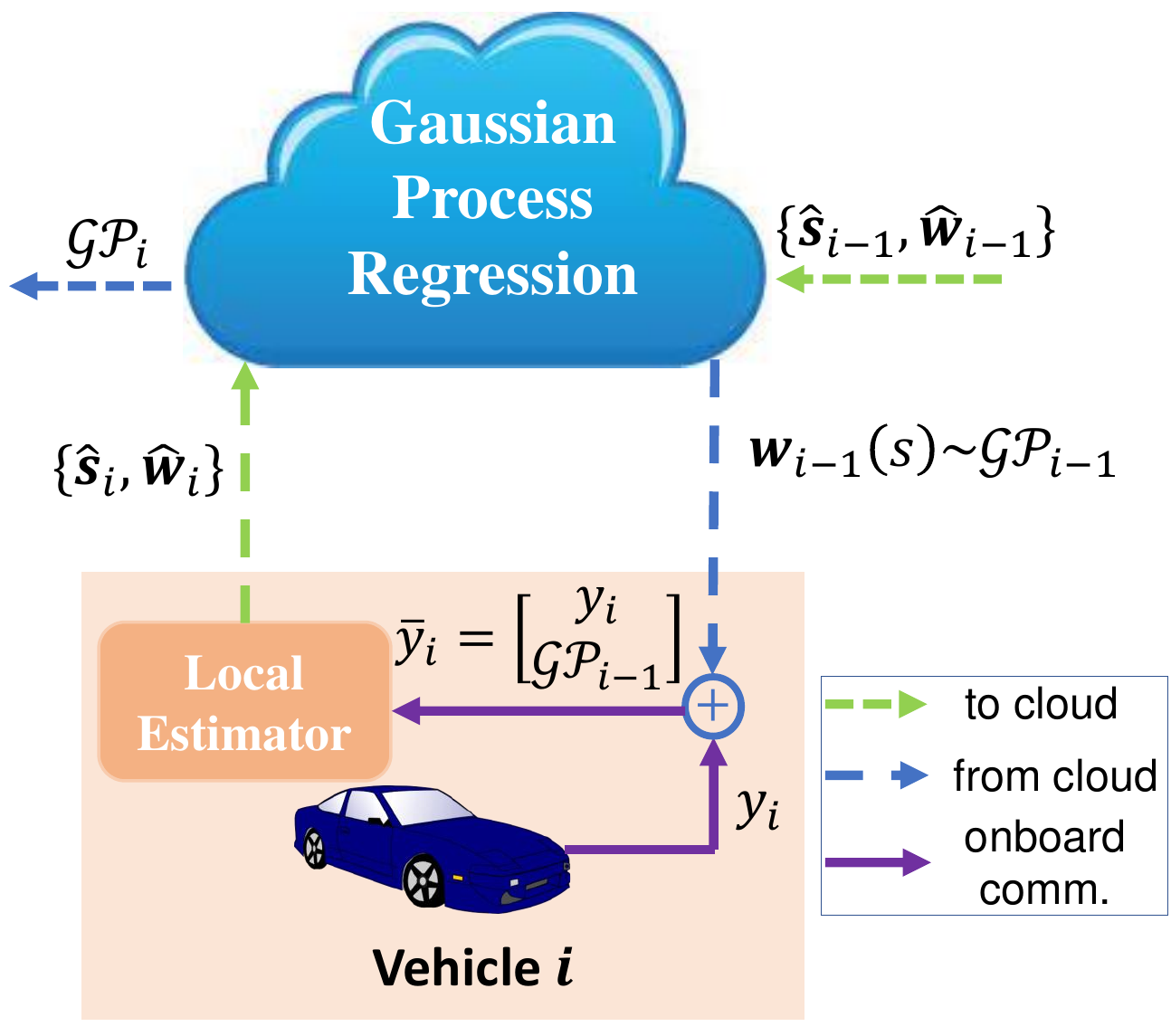}
    \caption{Schematics of cloud-assisted collaborative estimation using crowdsourced GP as pseudo-measurements.}
    \label{fig:GP}
\end{figure}

\section{Cloud-Assisted Collaborative Road information discovery with GP}\label{sec:coll}
In this section we detail our cloud-based collaborative estimation framework for road information discovery, systematically integrating a GP on the cloud side and an KF on the vehicle side. 
As illustrated in Fig.~\ref{fig:GP}, in this framework each participating vehicle $i$ receives from cloud a Gaussian process $\mathcal{GP}_{i-1}(m_{i-1}(s),\mathcal{K}_{i-1}(s,s')|\hat{\Theta}_{i-1})$ with mean function $m_{i-1}(s)$ and kernel function $\mathcal{K}_{i-1}(s,s')$, parameterized by $\hat{\Theta}_{i-1}$, which was trained on the aggregated data from prior participating vehicles until vehicle $i-1$ . The GP characterization of the road condition is advantageous as it describes both the estimate (by the mean) as well as the uncertainty (by the kernel function). The received $\mathcal{GP}_{i-1}(\cdot\,,\,\cdot\,|{\Theta}_{i-1})$ is then utilized as {\it a priori} ``pseudo-measurement'' of the road information parameter, forming an augmented output along with the onboard measurements. More specifically, at each time step $k$, an augmented output is formed as
 \begin{equation}
      \bar{\mathbf{y}}_i(k)=[\mathbf{y}_i(k);\, {{w}}_{gp,i-1}(\hat{s}(k))],
      \label{Eq:Augy}
 \end{equation}
where ${{w}}_{gp,i-1}(\hat{s}(k))=m_{i-1}(\hat{s}(k))$ is the mean function evaluated at $\hat{s}(k)$ with $\hat{s}(k)$ being the estimated vehicle position (e.g., from GPS) at time step $k$. This augmented output is then incorporated into a local KF estimator, to estimate the augmented state (both vehicle states and road information parameter). In this setting, the output matrix and the measurement noise covariance in KF (Eqn.~\ref{Eq:KF1}) after incorporating $\mathcal{GP}_{i-1}$  are modified as:
\begin{equation}
   {\Tilde{C}_i}=\left[ \begin{matrix}
   {} & \bar{{C}}_i & {}  \\
   0 & \dots & 1  \\
\end{matrix} \right],\, \bar{{R}}_i({k})=\left[ \begin{matrix}
   {R} & {0}  \\
   {0} & \text{Var}\left(w_{_{gp,i-1}}(k)\right)   \\
\end{matrix} \right],
\label{eq:aug2}
\end{equation}
where $w_{_{gp,i-1}}(k)$ is a short notation for $w_{_{gp,i-1}}(\hat{s}(k))$ and is adopted hereafter, and $\text{Var}\left(w_{_{gp,i-1}}(k)\right)$ denotes the variance of $w_{_{gp,i-1}}(k)$ and can be computed using the GP Kernel function as $\text{Var}\left(w_{_{gp,i-1}}(k)\right)=\mathcal{K}\left(\hat{s}(k),\hat{s}(k)\right)$. This essentially uses the cloud GP as a ``pseudo-measurement'' and we next show that this scheme can enhance the performance of KF by reducing the minimum mean square error (MMSE). 

\subsection{GP pseudo-measurement for MMSE reduction}\label{sec:pseudo}
In this subsection, we show that the GP pseudo-measurement strategy described above can improve the performance of KF by reducing the MMSE. In this analysis, we consider the discrete-time dynamic system (\ref{Eq:DT}), without subscript $i$ for notational simplicity. The performance of a state estimator can be characterized by the mean square error (MSE) defined as: 
\begin{equation}
\operatorname{MSE}\left(\hat{\bar{\mathbf{x}}}(k)\right) := \mathbb{E}\{(\bar{\mathbf{x}}(k)-\hat{\bar{\mathbf{x}}}(k))^{T}(\bar{\mathbf{x}}(k)-\hat{\bar{\mathbf{x}}}(k)\},
\end{equation}
where $\hat{\bar{\mathbf{x}}}(k)$ is the estimate from a state estimator and $\bar{\mathbf{x}}(k)$ is the true state. It is shown in \cite{kalman1960new} that assuming Gaussian process noise and measurement noise, the Kalman filter in Eqns.~\ref{Eq:KF1}-\ref{Eq:KF2} is optimal in the sense that it achieves the minimum 
MSE (MMSE), that is,
\begin{equation}
\operatorname{MSE}_{KF}\left(\hat{\bar{\mathbf{x}}}(k)\right) =\operatorname{MMSE}=tr(P(k|k)),
\end{equation}
where $P(k|k)$ is the state estimate covariance defined in (\ref{Eq:KF1}).

Let $\mathcal{S}_1$ denote the original sensor configuration with $r$ onboard measurements, corresponding to the measurement matrix $\bar{C}$ in (\ref{Eq:DT1}), and let $\mathcal{S}_2$ denote the sensor configuration using the GP as an additional pseudo-measurement, totaling $r+1$ measurements and corresponding to the $\tilde{C}$ in (\ref{eq:aug2}). Note that one can perform a change-of-state transformation such that the $\bar{C}$ matrix can be transformed to $\bar{C}_{r}$, which is a rank-$r$ matrix with $r$ rows (corresponding to $r$ measurements) and each row of $\bar{C}_{r}$ has exact one element equal to 1 and others are all 0 \cite{tzoumas2016sensor}.  
  The following proposition summarizes the benefits of the GP pseudo-measurement. 

\textbf{Proposition 1:} Consider the sensor configurations $\mathcal{S}_1$ and $\mathcal{S}_2$ defined above, with output matrices $\bar{C}_{{r}}$ and 
\begin{equation}
    \bar{C}_{r+1}= 
\begin{bmatrix}
    \bar{C}_{{r}}   \\
  l 
\end{bmatrix}, \quad l=[0\;0\cdots\;0\;1]\in\mathbb{R}^{1\times n},
\label{eq:c_(r+1)}
\end{equation}
respectively. In addition, assume the process noise $\bar{\eta}$ and measurement noise $v$ in (\ref{Eq:DT1}) are Gaussian with $\mathbb{E}\{vv^{T}\}=\sigma^{2}_{v}I$ and $\mathbb{E}\{\bar{\eta} \bar{\eta}^{T}\}=\sigma^{2}_{\bar{\eta}}I$. Then
\begin{equation}
   \text{MMSE}({\hat{\bar{{\mathbf{x}}}}_{r}}(k'))>\text{MMSE}({{\hat{{
\bar{\mathbf{{x}}}}}}_{{{r+1}}}}(k')).
\end{equation}
where $\hat{\bar{\mathbf{x}}}_{r}(k')$ and $\hat{\bar{\mathbf{x}}}_{r+1}(k')$ are, respectively, the KF estimates with sensor sets $\mathcal{S}_1$ and $\mathcal{S}_2$ at time step $k'$,  $k'\in [0,k]$.

\textit{Proof:} Following the notations in \cite{tzoumas2016sensor}, for the $\mathcal{S}_1$ sensor configuration Eqn.~\ref{Eq:DT1} implies that
\begin{equation}
  {{\mathbf{y}}_{1:k}}={{O}_{k,r}}{{z}_{k-1}}+{{v}_{1:k}},
  \label{eq:allys}
\end{equation}
where the subscript ${1:k}$ denotes the vector concatenation from step 1 to step $k$,  $z_{k-1}=[\bar{\mathbf{x}}_{0};\;\bar{\eta}_{1:k-1}]$, $O_{k,r}=[\bar{C}_rL_0;\;\bar{C}_rL_1;\;\cdots;\;\bar{C}_rL_k]$ with
\begin{equation}
    L_i = \begin{cases} [{{I}_{n\times n}}\quad{{0}_{n\times n}}] &\mbox{for } i=0, \\
[{{{\bar{A}}}^{i}}\quad {{{\bar{A}}}^{i-1}} \cdots \quad \bar{A}^i\quad I_{n\times n}] & \mbox{for } i=1,2,\ldots,k. \end{cases}
\end{equation}

Let's denote the covariance matrix of $\hat{z}_{k-1,r}$ with the first sensor configuration as $\Sigma_{z_{k-1,r}}=\mathbb{E}\left\{(z_{k-1}-\hat{z}_{k-1,r})(z_{k-1}-\hat{z}_{k-1,r})^T\right\}  $. From the maximum-likelihood principle \cite{mendel1995lessons} and Eqn.~\ref{eq:allys}, it follows that
\begin{equation}
\label{eqn:Sigma_z}
\begin{aligned}
\Sigma_{z_{k-1,r}}&=\sigma _{\bar{\eta}}^{2}I-\sigma _{\bar{\eta}}^{2}IO_{k,r}^{T}[\sigma _{\bar{\eta}}^{2}I{{O}_{k,r}}O_{k,r}^{T}+\sigma _{v}^{2}I]\sigma _{\bar{\eta}}^{2}I{{O}_{k,r}}\\&={{(\sigma _{\bar{\eta}}^{-2}I+\sigma _{v}^{-2}IO_{k,r}^{T}{{O}_{k,r}})}^{-1}},
\end{aligned}
\end{equation}
where the second line is from the Woodbury matrix identity \cite{bernstein2009matrix}.
Also  Eqn.~\ref{eq:c_(r+1)} implies that
\begin{equation}
\bar{C}_{r+1}^{T}{\bar{C}_{r+1}} =\bar{C}_{r}^{T}{\bar{C}_{r}}+ l^Tl\succ \bar{C}_{r}^{T}{\bar{C}_{r}},
\label{eq:c^transpose_c}
\end{equation}
from which and the fact that $rank(L_0)=rank(L_1)=\cdots= rank(L_k)=n$, according to Proposition 8.1.2 of \cite{bernstein2009matrix}, it follows that
\begin{equation}
\label{eqn:O_comp}
    \begin{aligned}
    &O_{k,r+1}^{T}{{O}_{k,r+1}}\\
    &=\left[{{L}_{0}}^{T}{\bar{C}_{r+1}^{T}}\quad\cdots\quad{{L}_{k}}^{T}{\bar{C}_{r+1}^{T}}\right] \left[\bar{C}_rL_0;\;\cdots;\;\bar{C}_rL_k\right]\\
  &={{L}_{0}}^{T}{\bar{C}_{r+1}^{T}}\bar{C}_{r+1}{{L}_{0}}+\dots
 +{{L}_{k}}^{T}{\bar{C}_{r+1}^{T}}\bar{C}_{r+1}{{L}_{k}} \\ 
 & \succ O_{k,r}^{T}{{O}_{k,r}}.
    \end{aligned}
\end{equation}
With the non-negativity of $\sigma_{\bar{\eta}}$ and $\sigma_{v}$, (\ref{eqn:O_comp}) implies that
\begin{equation}
\begin{aligned}
\sigma _{\bar{\eta}}^{-2}I+\sigma _{v}^{-2}O_{k,{{r+1}}}^{T}{{O}_{k,{{r+1}}}}\succ \sigma _{\bar{\eta}}^{-2}I+\sigma _{v}^{-2}O_{k,{{r}}}^{T}{{O}_{k,r}},
\end{aligned}
\end{equation}
or equivalently,
\begin{equation}
 {{(\sigma _{\bar{\eta}}^{-2}I+\sigma _{v}^{-2}O_{k,{r+1}}^{T}{{O}_{k,{{r+1}}}})}^{-1}}\prec {{(\sigma _{\bar{\eta}}^{-2}I+\sigma _{v}^{-2}O_{k,{{r}}}^{T}{{O}_{k,{{r}}}})}^{-1}},   
\end{equation}
which shows that $\Sigma_{z_{k-1},r} \succ \Sigma_{z_{k-1},r+1}$ according to (\ref{eqn:Sigma_z}). Recall that $\hat{\bar{\mathbf{x}}}(k')=L_{k'}\hat{z}_{k-1}$ for any $k' \in [0,k]$ and that MMSE$(\hat{\bar{\mathbf{x}}}(k'))=tr(L_{k'}\Sigma_{z_{k-1}}L_{k'}^{T})$. Therefore,  MMSE$({{\hat{\bar{\mathbf{x}}}}_{{{r}}}}(k'))>$MMSE$({{\hat{\bar{\mathbf{x}}}}_{{{r+1}}}}(k'))$ holds (see the trace inequality in  \cite{bernstein2009matrix} Corollary 8.4.10), which completes the proof. $\hfill \blacksquare$

\subsection{Fixed Interval Smoothing}
As the cloud-based crowdsourcing does not require real-time data uploading, if onboard memory permits, one can further improve the estimation performance through smoothing \cite{mendel1995lessons}. More specifically, consider a road segment that has been estimated by vehicle $i$ using the pseudo-measurement augmented KF discussed above and assume that  the estimates and the covariances are recorded onboard, the following smoothing step can be performed:
\begin{equation}
  \begin{aligned}
& \textbf{Smoothing:} \\ 
& {H}_i(k) \triangleq {P}_i(k| k) \bar{A}^{T}_i {P}_i^{-1}(k+1|k) \\
& \hat{\bar{\mathbf{x}}}_i(k|T_f)=\hat{\bar{\mathbf{x}}}_i(k|k)+{H}_i(k)[\hat{\bar{\mathbf{x}}}_i(k+1|T_f)-\hat{\bar{\mathbf{x}}}_i(k+1|k)],
\end{aligned}
\label{eq:smooth}
\end{equation}
where $T_f$ is the total number of estimation steps for the considered road segment.
This specific way of smoothing is called fixed-interval smoothing \cite{mendel1995lessons} and it is essentially a backward KF to make corrections on earlier estimates by incorporating later measurements.  This strategy can generally improve the estimation performance  \cite{lee2020adaptive} and is thus  adopted in our framework. 

\subsection{Cloud-based Crowdsourcing with GP}
With smoothed states from vehicle $i$, the road estimates along with the synchronized vehicle location is $(\hat{\mathbf{s}}_i,\hat{\mathbf{w}}_{s,i}):=\{\hat s_i(k),\hat w_i(k|T_f)\}_{k=1}^{T_f}$, where $\hat{s}(k)$ is the estimated/measured vehicle position at time step $k$, and $T_f$ again is the total number of estimation steps for the considered road segment. Gathering all the collected estimates up to vehicle $i$, a Gaussian process is trained on the cloud to crowdsource the collected data,
$\mathcal{D}_i=\{(\mathbf{\hat{s}_1},\mathbf{\hat{w }}_{s,1}),(\mathbf{\hat{s}}_2,\mathbf{\hat{w }}_{s,2}),...,({\mathbf{\hat{s}}_{i}},\mathbf{\hat{w}}_{s,i})\}$, 
which is used to update the GP hyper-parameters defined in Section~\ref{sec:GP} using e.g., maximum likelihood learning:
\begin{equation}\label{equ:MLE}
\begin{aligned}
    \hat{\Theta}_i&=\argmax_\Theta \text{Prob}\Large[\large(\hat{\mathbf{w}}_{s,1}(\hat{\mathbf{s}}_1),\hat{\mathbf{w}}_{s,2}(\hat{\mathbf{s}}_2),\cdots,\hat{\mathbf{w}}_{s,i}(\hat{\mathbf{s}}_i)\large)|\hat{\mathbf{s}}_1,\\
    &\qquad\hat{\mathbf{s}}_2,\cdots,\hat{\mathbf{s}}_i,\Theta\Large].
    \end{aligned}
\end{equation}

 In this regard, one can stack the training data from multiple vehicles by stacking them as $\mathbf{\hat{S}}=[\mathbf{\hat{s}}_{1},\mathbf{\hat{s}}_{2}, ..., \mathbf{\hat{s}}_{i}]$ and $\mathbf{\hat{{W}}}=[\mathbf{\hat{w }}_{s,1},\mathbf{\hat{w }}_{s,2}, ..., \mathbf{\hat{w }}_{s,i}]$ respectively. The objective is to approximate the spatial road information distribution as nonlinear mapping:
\begin{equation}
   w=f({{s}_{}})+\epsilon_{w},
\end{equation}
with additive white Gaussian noise ${\epsilon_w }\sim \mathcal{N} ({0},\sigma^{2}_{{w}})$. For the GP prior, it follows that the output data is a related normal distribution
\begin{equation}
    \mathbf{\hat{W}}\sim \mathcal{N}({0},\bm{\mathcal{K}}(\mathbf{\hat{S}},\mathbf{\hat{S}})+\sigma^{2}_{{w}}{I}),
\end{equation}
 where $\bm{\mathcal{K}}(\mathbf{\hat{S}},\mathbf{\hat{S}})$ is the covariance matrix for input data. With the tuned hyperparameters of the kernel and mean function one can make predictions by posterior inference conditional on observed data $\mathcal{D}_i$. 
 Using these information, the predictive equations for the $i$'th GP regression at points $\mathbf{s}_{*}$  follow as
\begin{equation}
   { \mathbf{w}}_{_{gp,{i}}}=\mathbb{E}[\mathbf{w}_{*}|\mathbf{\hat{S}},\mathbf{\hat{W}} ,{{\mathbf{s}}_{*}}]=\bm{\mathcal{K}}({{\mathbf{s}}_{*}},\mathbf{\hat{S}}){{[\bm{\mathcal{K}}(\mathbf{\hat{S}},\mathbf{\hat{S}})+{{\sigma^{2}_{{w}}{I}}}}]^{-1}}\mathbf{\hat{W}}, 
    \label{Eq:mean}
\end{equation}
\begin{equation}
\label{Eq:cov}
\begin{aligned}
\text{Cov}({\mathbf{w}}_{_{gp,i}})&=\bm{\mathcal{K}}({\mathbf{s}}_{*},{\mathbf{s}}_{*})-\bm{\mathcal{K}}({{\mathbf{s}}_{*}},\mathbf{\hat{S}})\times \\ & {{[\bm{\mathcal{K}}(\mathbf{\hat{S}},\mathbf{\hat{S}})+{{\sigma^{2}_{{w}}{I}}}}]^{-1}}\bm{\mathcal{K}}({\mathbf{\hat{S}}},{{\mathbf{s}}}_{*}).
\end{aligned}
\end{equation}

The updated GP with hyper-parameters $\hat\Theta_i$, $\mathcal{GP}_i(\cdot\,,\,\cdot\,|\hat{\Theta}_{i})$, is then sent to the next participating vehicle $i+1$ to enhance its estimate as described in Section~\ref{sec:pseudo}. The process is then repeated.

\subsection{Noisy-Input GP to Handle GPS Uncertainty}
In practice, the position estimate $\hat{s}_{i}$ is usually corrupted with nonnegligible noises as the Global Positioning System (GPS) used in everyday vehicles has limited resolution. Hence, the standard GP regression discussed above, which does not account for noisy inputs, will inevitably increase the prediction variance. To obviate this issue, we explicitly incorporate the GPS uncertainty and employ a Noisy Input Gaussian Process or NIGP regression  \cite{mchutchon2011gaussian} to enhance the performance. The method follows from the idea that the influence of the input noise is proportional to the gradient of the input-output mapping.
Specifically, for each training point, we fit a local linear model to the GP posterior mean; the output can then be referred to by the input noise variance, which is proportional to the square of the gradient of the posterior mean function. In particular, we consider that the GPS measurement is corrupted by an additive noise, i.e.,
\begin{equation}
    \hat{s}=s+\epsilon_{s},
\end{equation}
where $\epsilon_{s} \sim \mathcal{N}(0,\sigma^2_{s})$, and the output can thus be represented as
\begin{equation}
    w=f(s+\epsilon_{s})+\epsilon_w.
    \label{eq:o_withnoisy_I}
\end{equation}
Following \cite{mchutchon2011gaussian}, one can approximate the input-output relationship using the first-order Taylor expansion as:
\begin{equation}
    w\approx f(s)+\epsilon_{s}f'(s) +\epsilon_{w},
    \label{eq:o_withTaylor}
\end{equation}
where $w'(s)=\frac{dw}{ds}$. Eqn.~\ref{eq:o_withTaylor} shows that the contribution of the input noise to the output noise is approximately proportional to the function gradient evaluated at the input, which  results in a prior observation distribution as:
\begin{equation}
    w \sim \mathcal{N}(0,\sigma^2_{w}+(w'(s)\sigma_{s})^2).
    \label{eq:prior_y}
\end{equation}
In addition, the GP prediction can then be calculated as: 
\begin{equation}
\begin{aligned}
\mathbf{w}_{nigp,i}&=\mathbb{E}[\mathbf{w}_{*}|
\mathbf{\hat{S}},\mathbf{\hat{W}} ,\mathbf{s}_{*}]=\bm{\mathcal{K}}(\mathbf{s}_{*},\mathbf{\hat{S}})[\bm{\mathcal{K}}(\mathbf{\hat{S}},\mathbf{\hat{S}})+\sigma^{2}_{w}I \\ & 
+(w'(s)\sigma_{s})^2{I}]^{-1}\mathbf{\hat{W}}, 
\end{aligned}
\label{Eq:NIGPmean}
\end{equation}
\begin{equation}
\label{Eq:NIGPcov}
\begin{aligned}
{cov}(\mathbf{w}_{_{nigp,i}})=&\bm{\mathcal{K}}({\mathbf{s}}_{*},{\mathbf{s}}_{*})-\bm{\mathcal{K}}({{\mathbf{s}}_{*}},\mathbf{\hat{S}})\times \\ & {{[\bm{\mathcal{K}}(\mathbf{\hat{S}},\mathbf{\hat{S}})+{{\sigma^{2}_{{w}}{I}}}}+(w'(s)\sigma_{s})^2{I}]^{-1}}\bm{\mathcal{K}}({\mathbf{\hat{S}}},{{\mathbf{s}}}_{*}).
\end{aligned}
\end{equation}
The input data noises are thus treated deterministically with a corrective term, $(w'(s)\sigma_{s})^2\mathbf{I}$, added to the output noise. In addition, one extra hyperparameter is added to the hyperparameter set, i.e. $\sigma_s$, which characterizes the GPS uncertainty. More details on how the NIGP model is used for training and prediction can be found in \cite{mchutchon2011gaussian}. We show in Section~\ref{sec:app} that, as compared to the standard GP, the NIGP approach can reduce estimation variance.

\begin{remark}
The proposed cloud-assisted collaborative estimation framework has the following advantages. First, it works for a fleet of heterogeneous vehicles as the framework has no requirement in vehicle homogeneity; each vehicle exploits its own model  for local estimation.
Second, the GP-based ``pseudo-measurement'' scheme is guaranteed to enhance the local onboard estimation performance as shown in Proposition 1 \textcolor{black}{(Section III-A)}.
Third, as only information regarding road estimate is sent to the cloud, privacy-sensitive information such as vehicle states are inherently protected. The proposed collaborative estimation paradigm combines the local measurements from individual/heterogeneous vehicles equipped with sensors of limited accuracy to achieve iteratively enhanced estimation performance (see results in later sections) and can explicitly deal with GPS uncertainties, making it practically appealing. 
\end{remark}

\section{Application to road profile estimation}\label{sec:app}
In this section, we apply the cloud-assisted collaborative estimation framework to an important yet challenging application: road profile estimation, the exploitation of which  has recently received significant interests in vehicle controls to improve safety and comfort \cite{Preview1,Preview2,Preview3}. Specifically, a road profile characterizes the detailed road elevation and can be characterized by $w(s)$, a function of the spatial distance as in Fig.~\ref{Eq:DT}. We demonstrate the efficacy of the proposed framework in the application of road profile estimation on both simulations and hardware-in-the-loop experiments.
\subsection{System Dynamics}
\begin{figure}[!t]
    \centering
    \includegraphics[width=0.65\linewidth]{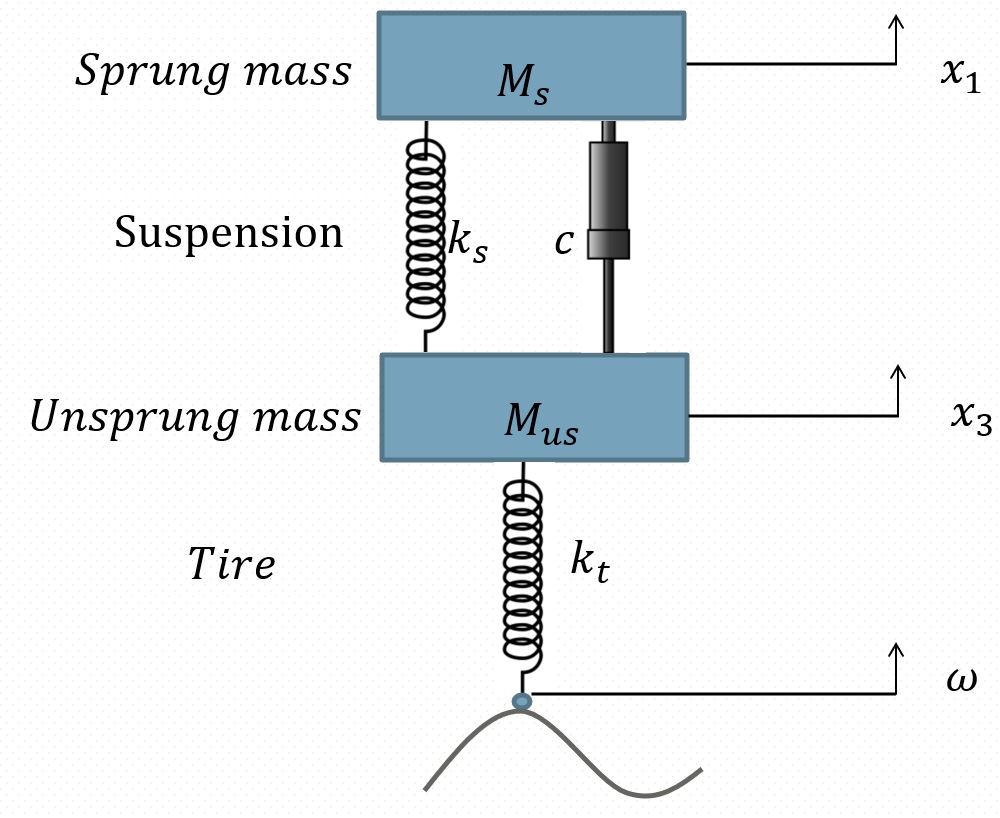}
    \caption{Illustration of a quarter-car suspension model.}
    \label{fig:quartercar}
\end{figure}

A quarter-car suspension model, as illustrated in Fig.~\ref{fig:quartercar}, is adopted here to characterize the vehicle-road interaction, and the general system dynamics in (\ref{Eq:DT}) now takes the following specific form  (we drop the index $i$ here to simplify the notation):
\begin{equation}\label{eqn:quarter}
\begin{aligned}
 & {{{\dot{x}}}_{1}}={{x}_{2}}, \\ 
 & {{{\dot{x}}}_{2}}=\frac{1}{{{M}_{s}}}(-k_s{{x}_{1}}-c{{x}_{2}}+k_s{{x}_{3}}+c{{x}_{4}}), \\ 
 & {{{\dot{x}}}_{3}}={{x}_{4}}, \\ 
 & {{{\dot{x}}}_{4}}=\frac{1}{{{M}_{us}}}(k_s{{x}_{1}}+c{{x}_{2}}-(k_s+{{k}_{t}}){{x}_{3}}-c{{x}_{4}}+{{k}_{t}}w), \\ 
 & y_1=x_1,\\
 & y_2=x_1-x_3,
\end{aligned}
\end{equation}
where $x_1$, $x_2$, $x_3$, and $x_4$  represent the sprung mass displacement, sprung mass velocity, unsprung mass displacement, and sprung mass velocity, respectively.
Here $w \in \mathbb{R} $ is the road
profile to be estimated, which can be modeled as the output of a low-pass filter driven by a white Gaussian noise with unit intensity. The low-pass filter transfer function is chosen as \cite{zuo2013energy}:
\begin{equation}
    G(s)=\frac{\sqrt{2\pi G_{r}V}}{s+w_0},
    \label{eq:filter}
\end{equation}
where $G_r$ is the road-roughness coefficient, $V$ is the vehicle's speed, and $w_0$ is the cut-off frequency. Alternatively, the low-pass filter can be written into a state-space form as $\dot{w}(t)=aw(t) +be(t)$, where $a=-w_0$, $b=\sqrt{2\pi G_{r}V}$, and $e$ is a white Gaussian noise with unit intensity.

{\color{black}{
\begin{remark}
Note that based on the road profile model (\ref{eq:filter}), the impact to the vehicle vertical dynamics due to the road  disturbance is a function of vehicle speed (proportional to the square root of the speed). As vehicles may have different travel speed, from (\ref{eq:filter}) one can use the scaling factor $\sqrt{\frac{V}{V_o}}$ to scale the local estimate to train the cloud-based GP. Similar speed-based scaling can be performed when downloading GP as pseudo-measurements for local estimation. Essentially, we train a cloud-based GP based on the road profile from a nominal speed (e.g., speed limit) and then scale it based on the actual vehicle speed. Alternatively, we can train several GPs for different speed ranges, e.g., 40-50mph, 50-60mph, etc. Based on the vehicle speed, we can then integrate the local estimation with an appropriate GP based on the travel speed. 
\end{remark}}

}

By augmenting the road profile $w$ as an additional state, i.e., ${\bar{\mathbf{{x}}}}=[x_1;x_2;x_3;x_4;w]$, and discretizing the system with an appropriate sampling time we obtain the  discrete time state-space equations in the form as Eqn.~\ref{Eq:DT1}, which can now be used in the  cloud-assisted collaborative estimation algorithm described in Section~\ref{sec:coll}.  Those steps are summarized in  Procedure~\ref{algo:1}.
In the following subsections, we will  show in simulations and experiments the efficacy of the developed framework. 

\begin{figure}[!t]
\removelatexerror
\scalebox{1}{
\begin{Procedure}[H]
\SetAlFnt{\small}

\caption{Cloud-based collaborative road profile estimation framework }
\label{algo:1}


$(\hat{\mathbf{s}}_1,\hat{\mathbf{w}}_1) \leftarrow$ Forward KF for vehicle 1;  \\

$\hat{\mathbf{w}}_{s,1}=\{\hat{{w}}_1(k|T_f)\}_{k=1}^{T_f} \leftarrow$ Backward smoothing for vehicle 1;  \\
 $({\mathbf{w}}_{_{{nigp,1}}},\text{Var}({\mathbf{w}_{_{nigp,1}}}))$ $\leftarrow$ NIGP $(\hat{\mathbf{s}}_1,\hat{\mathbf{w}}_1$); \\
 \For{$ i=2:N$}{
 $\mathcal{D}_{i-1}=\{({\mathbf{\hat{s}}_{1}},\mathbf{\hat{w}}_{s,1}),...,({\mathbf{\hat{s}}_{i-1}},\mathbf{\hat{w}}_{s,i-1})\}$;
 $({\mathbf{w}}_{_{{nigp,i-1}}},\text{Var}({\mathbf{w}_{_{nigp,i-1}}}))$ $\leftarrow$ NIGP $(\mathcal{D}_{i-1})$;

  $(\hat{\mathbf{s}}_i,\hat{\mathbf{w}}_i(k|k)) \leftarrow$ KF$_i ({\mathbf{w}}_{_{{nigp,i-1}}},\text{Var}({\mathbf{w}_{_{nigp,i-1}}}))$ \\
  $\hat{\mathbf{w}}_{s,i}=\{\hat{{w}}_i(k|T_f)\}_{k=1}^{T_f} \leftarrow$ Backward smoothing for vehicle $i$; \\
  $i=i+1;$
   }
\end{Procedure}
}
\end{figure}

\subsection{Simulation Setup}
In this section, simulation results for the proposed collaborative estimation framework are presented. Specifically, we consider $N=10$ heterogeneous vehicles with  different model parameters. The parameters used corresponding to Eqn.~\ref{eqn:quarter} for each vehicle $i=1,\dots,N$ are as Table \ref{paramet}.

\begin{table}[!h]\caption{Simulation parameters}
\begin{center}
\begin{tabular}{|c|c|c|c|}
\hline
\hline 
 $M_{s,i}\;[kg]$ & $M_{us}\;[kg]$ & $k_{t}\;  [kN/m]$ & $c\;[{Ns}/{m}]$ \\
\hline
$300\times(\frac{90+i}{100})$ & $60 $ & $190$ & $1000 $\\
\hline
 $k_{s,i}\;[kN/m]$ & $a$ & $b$ & $L$ \\
\hline
$16\times(\frac{90+i}{100})$ & $-0.01$ & $0.0328$ & $25$ \\
\hline
\hline
\end{tabular}
\end{center}

\label{paramet}
\end{table}

Two sensors, including sprung mass displacement and suspension \textcolor{black}{deflection}, are used, which corresponds to the following 
 ${C}$ matrix: $${C}= \begin{bmatrix}
   1 & 0 & 0 & 0 & 0  \\
   1 & 0 & -1 & 0 & 0  \\
\end{bmatrix}.$$

We consider a road segment of about 40-meters long, corresponding to a time span of 1.5 seconds with sampling time of $T_s=0.01s$, which results in a total of 151 estimation points for each KF.  The default GPS noise  is chosen as $\sigma_s=0.2$.   The measurement noise ${v}_{i}$ for each vehicle is generated in a way that the signal-to-noise (SNR) ratios are between 10 and 20, indicating relatively noisy sensors. \textcolor{black}{Also we add model uncertainties such as process and measurement noises that are unknown to the KF.}
 For the smoothing step, since the fixed interval smoothing incur higher computational cost as the number of estimated points and their dimension increase, we use a fixed-lag smoother type instead \cite{mendel1995lessons}, i.e., we condition the  estimated states at each time step $k$ on all the measurements up to time $L+k$ where $L$ is a fixed constant. As a result, we use $\hat{{w}}_i(k|k+L)=\mathbb{E}\{w_i(k)|\mathbf{Y}_i(k+L)\}$ for GP learning on the cloud. With $L$ being reasonably large, this fixed-interval strategy typically causes little performance degradation \cite{mendel1995lessons}, which is advantages for automotive applications with limited onboard resources.

 For the cloud-based NIGP, the initial prior is defined as a zero mean function with exponentiated quadratic kernel defined in Section~\ref{sec:GP}. 
For the NIGP regression, there are several possible approaches to calculate  Eqns.~\ref{Eq:mean} and \ref{Eq:cov}. The first approach is that, for the $i$th NIGP we use all the received data up to vehicle number $i$ for training and then infer the posterior given all the collected data. Another approach is to similarly collect all the data up to vehicle $i$ but  use a sparsity approximation \cite{snelson2007local} instead. Furthermore, one can adopt a recursive update scheme (see e.g., \cite{huber2014recursive}) where for the new arriving data the NIGP reuses past estimates for efficient updates. While data and computation efficient, this method can sometimes lead to performance degradation.  In this study,  we use the first approach as it is straightforward to implement and the data size in our study is manageable since the number of considered vehicles is relatively small.

\subsection{Simulation Results}
\subsubsection{Onboard Estimation Performance}
\begin{figure}[!h]
    \centering
    \includegraphics[width=0.8 \linewidth]{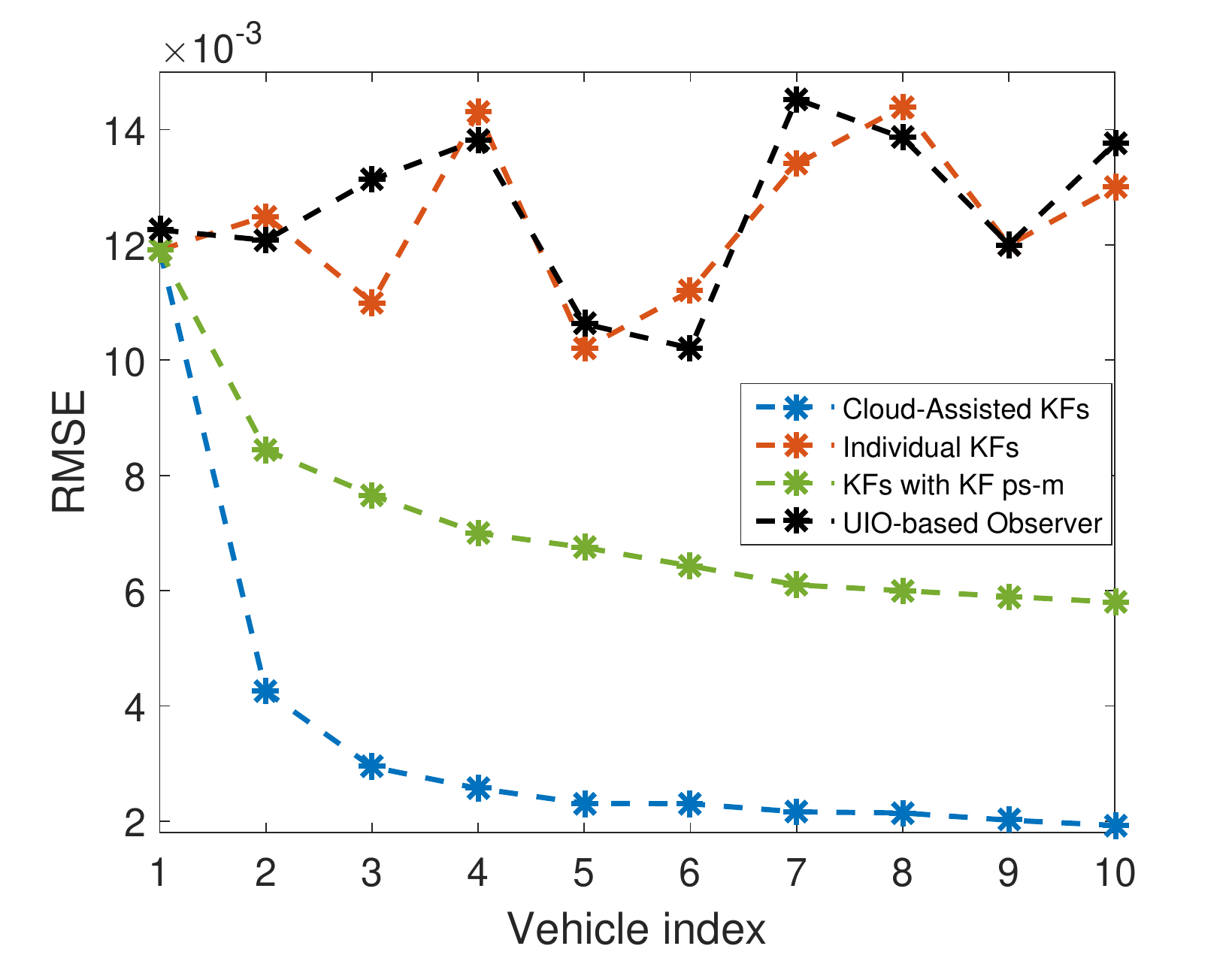}
    \caption{\textcolor{black}{Comparison of different onboard KF estimation schemes in RMSE.}}
    \label{fig:KF-PMrmse}
\end{figure}

\begin{figure}[!h]
    \centering
    \includegraphics[width=0.8 \linewidth]{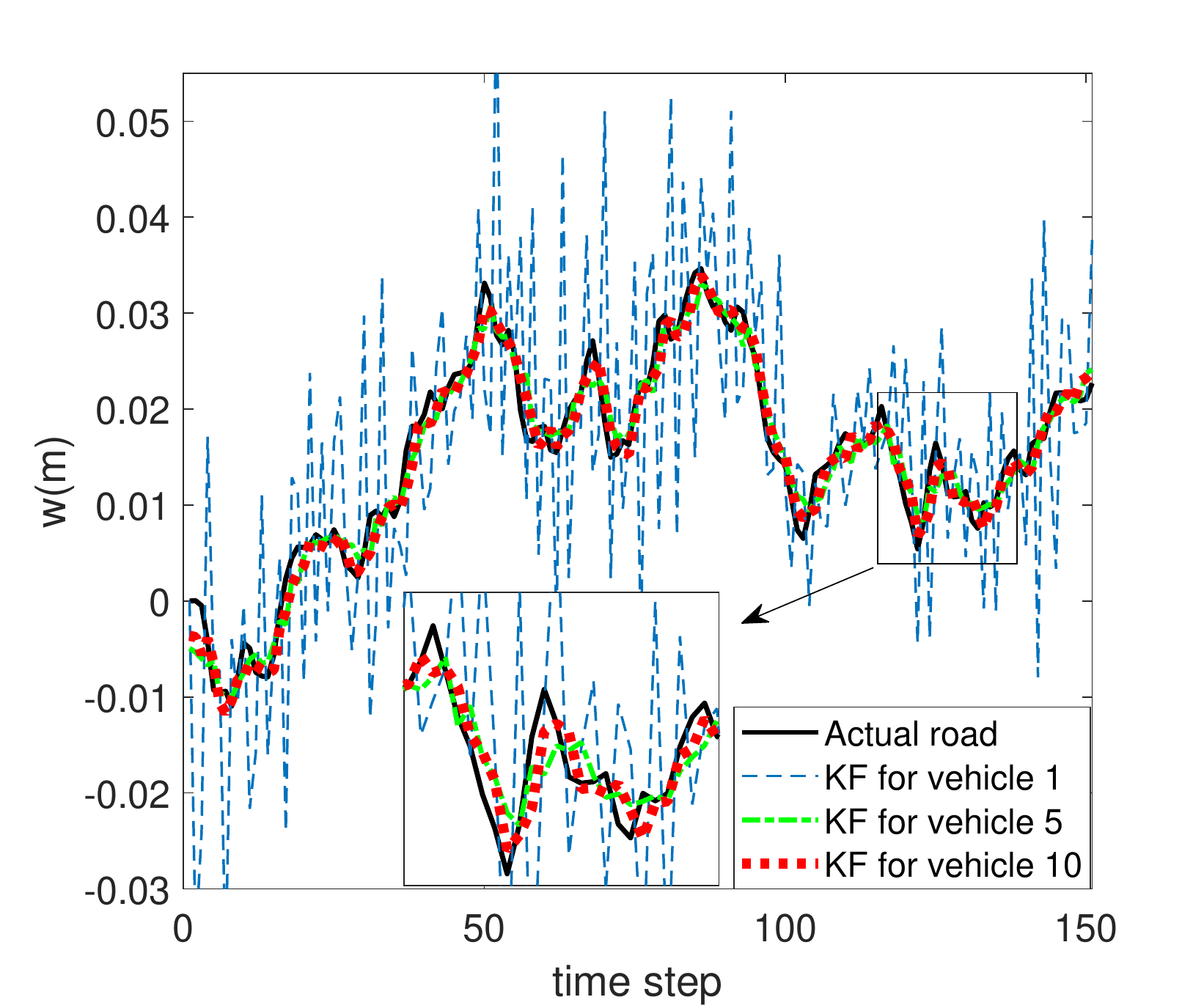}
    \caption{ Comparison of the resultant onboard estimation for the first, fifth, and last vehicle with the proposed framework to show the iterative onboard estimation improvement.}
    \label{fig:KF1_10}
\end{figure}

\begin{figure}[!h]
    \centering
    \includegraphics[width=0.8\linewidth]{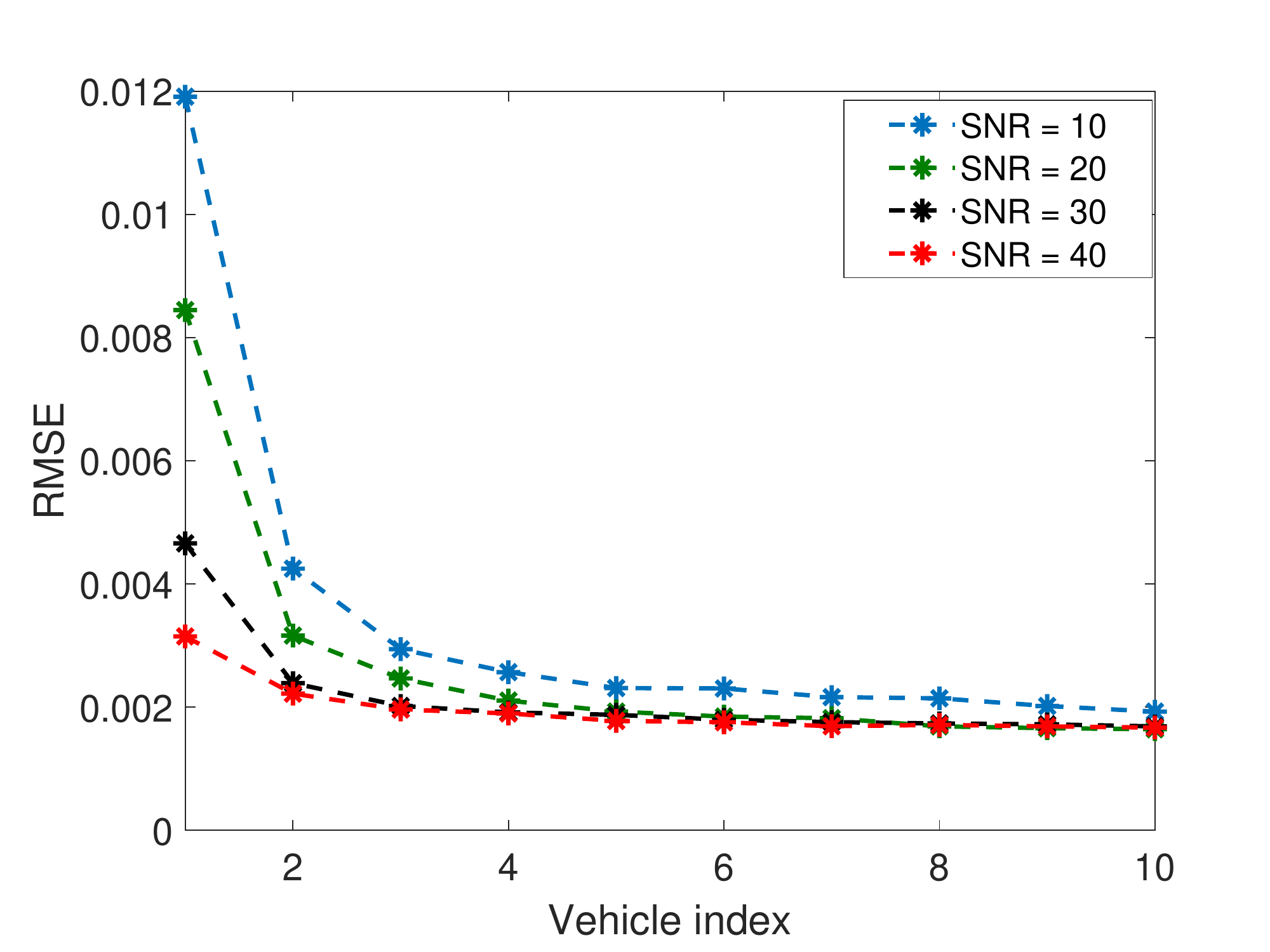}
    \caption{\textcolor{black}{The estimation performance under different signal-to-noise levels. It can be seen that the proposed framework can iteratively improve estimation performance despite noisy measurements.}}
    \label{fig:KFs_diffSNRs}
\end{figure}

Figs.~\ref{fig:KF-PMrmse}-\ref{fig:KFs_diffSNRs} summarize the performance of the proposed augmented KF  when  exploiting the latest NIGP fit as pseudo-measurements. In particular, Fig.~\ref{fig:KF-PMrmse} shows the root mean squared error (RMSE) between the actual road and each vehicle's estimation with onboard KF that uses NIGP pseudo-measurement, as well as its comparison with \textcolor{black}{two benchmark settings. In the first one, only onboard sensor measurements are used without exploiting the GP pseudo-measurements.The second benchmark is the implementation of the approach presented in \cite{kalabic2013multi} which is an UIO method.} We also include the performance when each vehicle uses a more simplified type of pseudo-measurements: incorporating the prior vehicle's KF estimation as an additional measurement (KFs with KF ps-m in Fig.~\ref{fig:KF-PMrmse}). 

It can be seen in Fig.~\ref{fig:KF-PMrmse} that for vehicles that only use onboard measurements the RMSEs randomly fluctuate due to the randomness in sensor measurements.
On the other hand,  employing the pseudo-measurement from the immediately preceding vehicles can greatly enhance the performance and show iterative improvement. Finally, the use of NIGP pseudo-measurement achieves the best performance as the crowdsourced NIGP provides more stable and reliable pseudo-measurements as compared to the use of KF from a preceding vehicle.
Fig.~\ref{fig:KF1_10} shows the iterative improvement on the onboard measurements when using the NIGP pseudo-measurement scheme, where  the first vehicle doesn't use any pseudo-measurement. It clearly demonstrates that the pseudo-measurement strategy can iteratively enhance onboard performance. \textcolor{black}{Finally, to better show the impact of measurement noise and the capability of handling large disturbances, we performed analysis of estimation results under different SNR levels, the results of which are summarized in Figure \ref{fig:KFs_diffSNRs}. It can be seen that our collaborative estimation scheme is able to utilize noisy measurements from multiple heterogeneous vehicles and iteratively refine the estimation performance. This shows the practical viability of our framework as the existing sensors in everyday vehicles are generally noisy. }

\subsubsection{Cloud NIGP Performance}
\begin{figure}[!h]
    \centering
    \includegraphics[width=0.9 \linewidth]{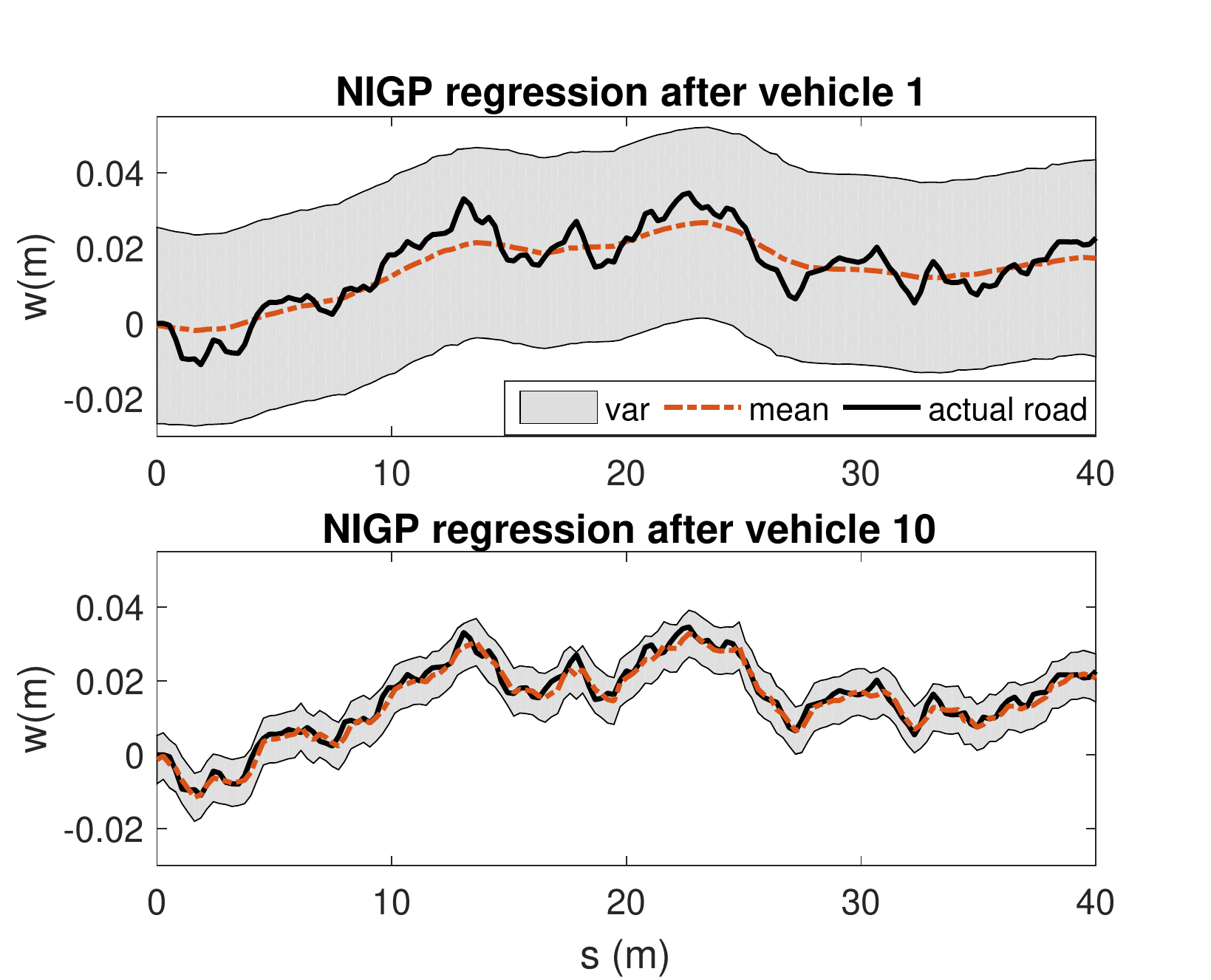}
    \caption{ NIGP regression in the cloud with one vehicle vs. with ten vehicles.}
    \label{fig:nigp1_10}
\end{figure}

\begin{figure}[!h]
    \centering
    \includegraphics[width=1 \linewidth]{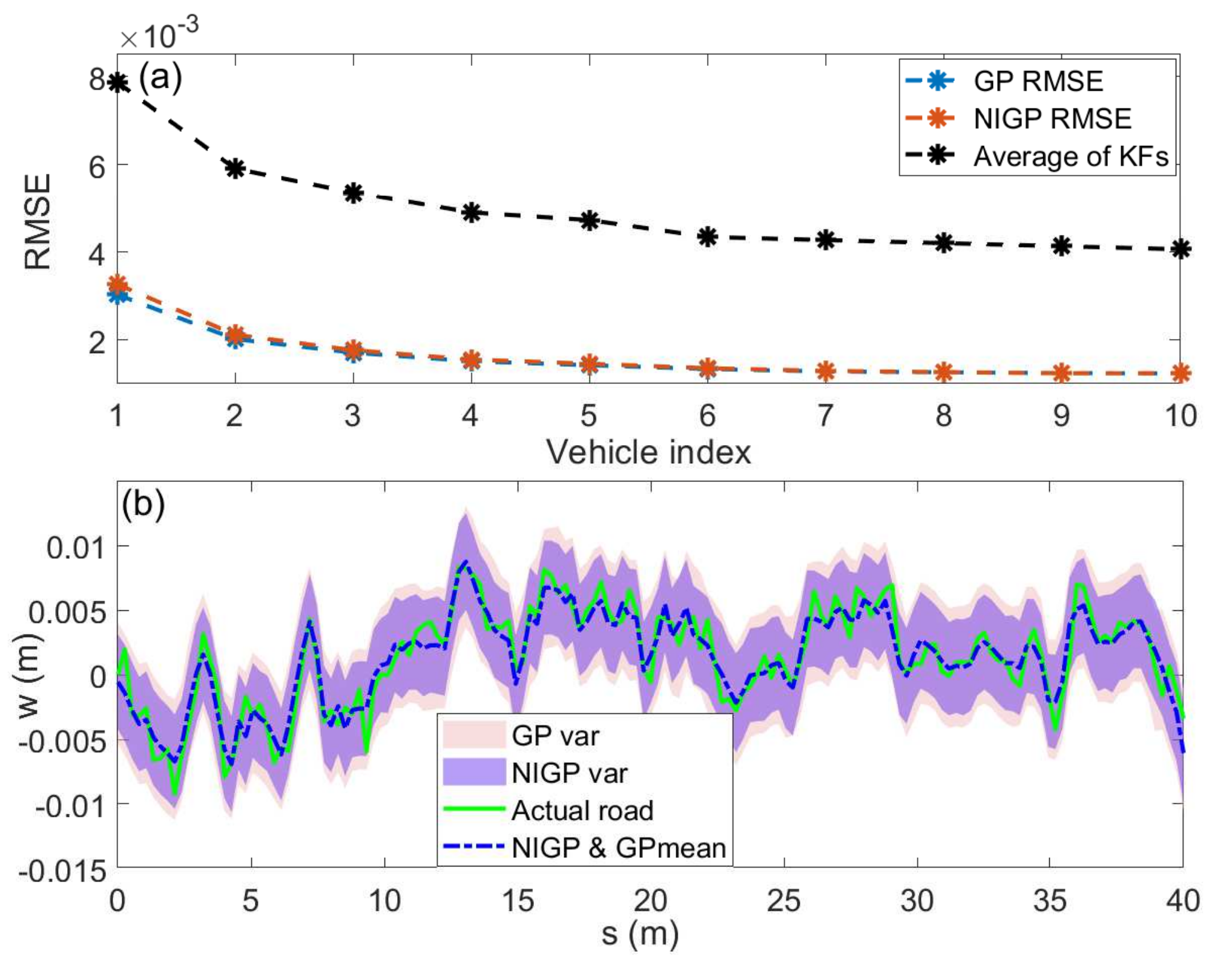}
    \caption{\textcolor{black}{(a): RMSE comparison of NIGP, GP and averaged KF. (b): Comparison of NIGP and GP fit with data from all ten vehicles. Only one mean function is plotted as the mean functions are almost identical in GP and NIGP.}}
    \label{fig:nigp_gp_and_cloudSim}
    \vspace{0pt}
\end{figure}

The performances of cloud-based GP  are summarized in Figs. \ref{fig:nigp1_10}-\ref{fig:nigp_gp_and_cloudSim}. 
It is clear in Fig. \ref{fig:nigp1_10} that the NIGP trained only on the first vehicle's data produces a model with poor road characterization while with the data from all ten vehicles, the road profile is well captured by the GP with much reduced variance. Fig.~\ref{fig:nigp_gp_and_cloudSim} (a) shows the RMSE comparison among the NIGP regression, standard GP regression, and a benchmark case where the KF estimations are fused through simple averaging. It can be seen that the GP and NIGP both significantly outperform the averaging benchmark.

While GP and NIGP result in similar RMSE, a clear difference in variance is shown in Fig.~\ref{fig:nigp_gp_and_cloudSim} (b), that is, NIGP produces estimates with reduced uncertainty. This is because NIGP regression explicitly accounts for the input noise  that effectively reduces the variance. In fact, the estimated variance after the last regression is $\hat{\sigma}_s=0.211$, which is very close to the actual value of $0.2$. 


\subsection{Hardware-in-the-loop Experiments}
\begin{figure}[!h]
    \centering
    \includegraphics[width=.65 \linewidth]{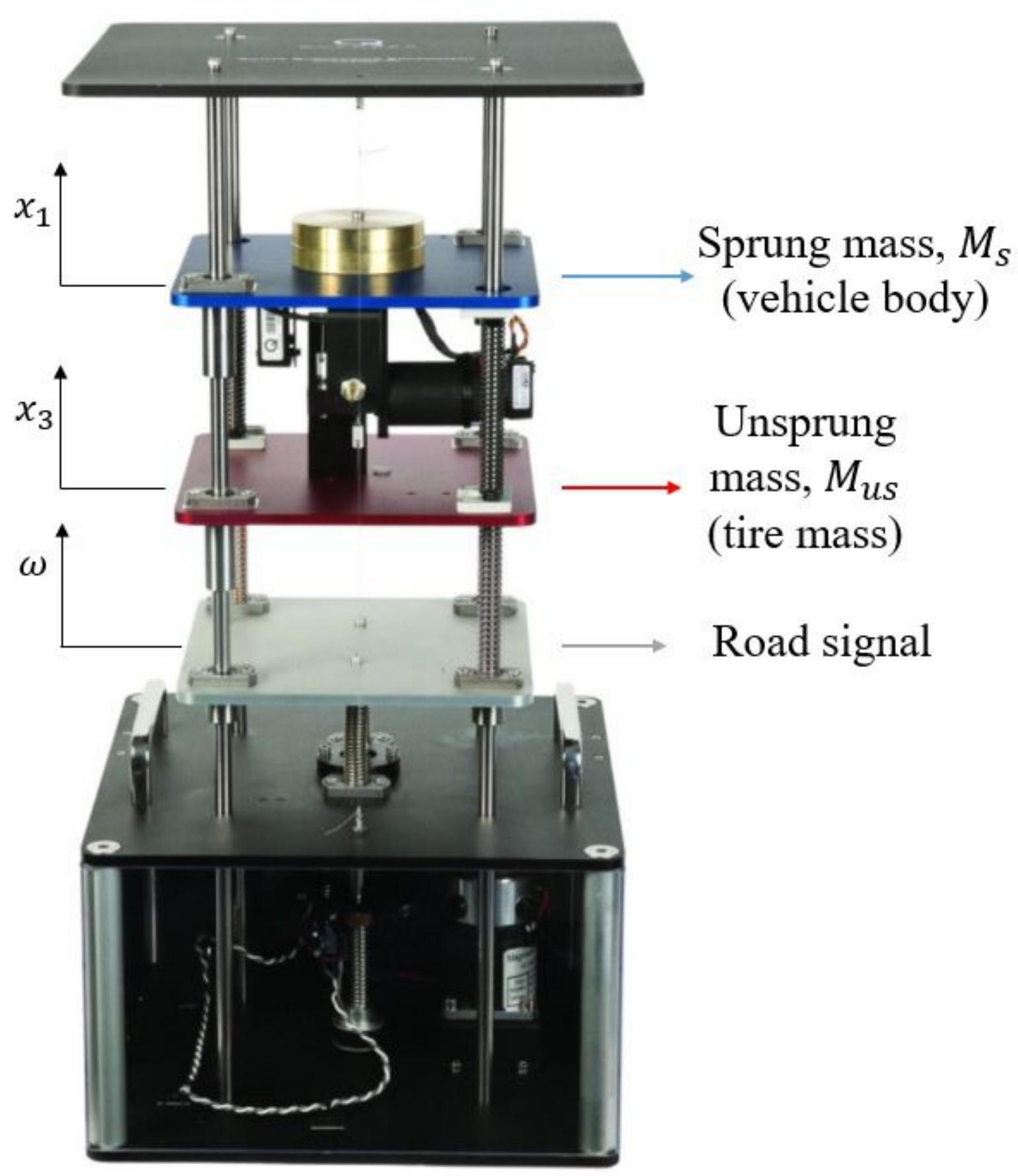}
    \caption{Quanser active suspension station used for experiments.}
    \label{fig:AS}
\end{figure}

We further implement our collaborative estimation framework on the Quanser Active Suspension (AS) platform (see Fig.~\ref{fig:AS}) to perform hardware-in-the-loop experiments. The suspension station resembles a quarter car in a smaller scale, and it is equipped with an actuator that can be used for active suspension controls. In this study,  we keep it passive (i.e., with zero controls) for the purpose of road profile estimation, which leads to the same dynamics as (\ref{eqn:quarter}).

\begin{table}[!h]
\caption{Parameters used in experiments (first run)}
\begin{center}
\begin{tabular}{|c|c|c|c|c|}
\hline
\hline 
 $M_{s,1}\; [kg]$ & $M_{us,1}\;[kg]$ & $k_{s,1}\; [{N}/{m}]$  & $a$ & $b$\\
\hline
$2.12 $ & $0.97$ & $999.99 $ & $-5$ & $0.0134$\\
\hline
$k_{t,1}\;[{N}/{m}]$  & $c_{t,1}\;[{Ns}/{m}]$ &  $c_{s,1}\;[{Ns}/{m}]$ &$L$ &\\
\hline
$1163.6$  & $7 $ & $9.5 $ & $25$ &\\
\hline
\hline
\end{tabular}
\end{center}\vspace{-3pt}
\label{paramet_exp}
\end{table}
The suspension station is equipped with encoders to measure the sprung mass displacement and suspension displacement in agreement with the $C$ matrix used in the simulation section. 

\begin{figure}[h]
    \centering
    \includegraphics[width=1 \linewidth]{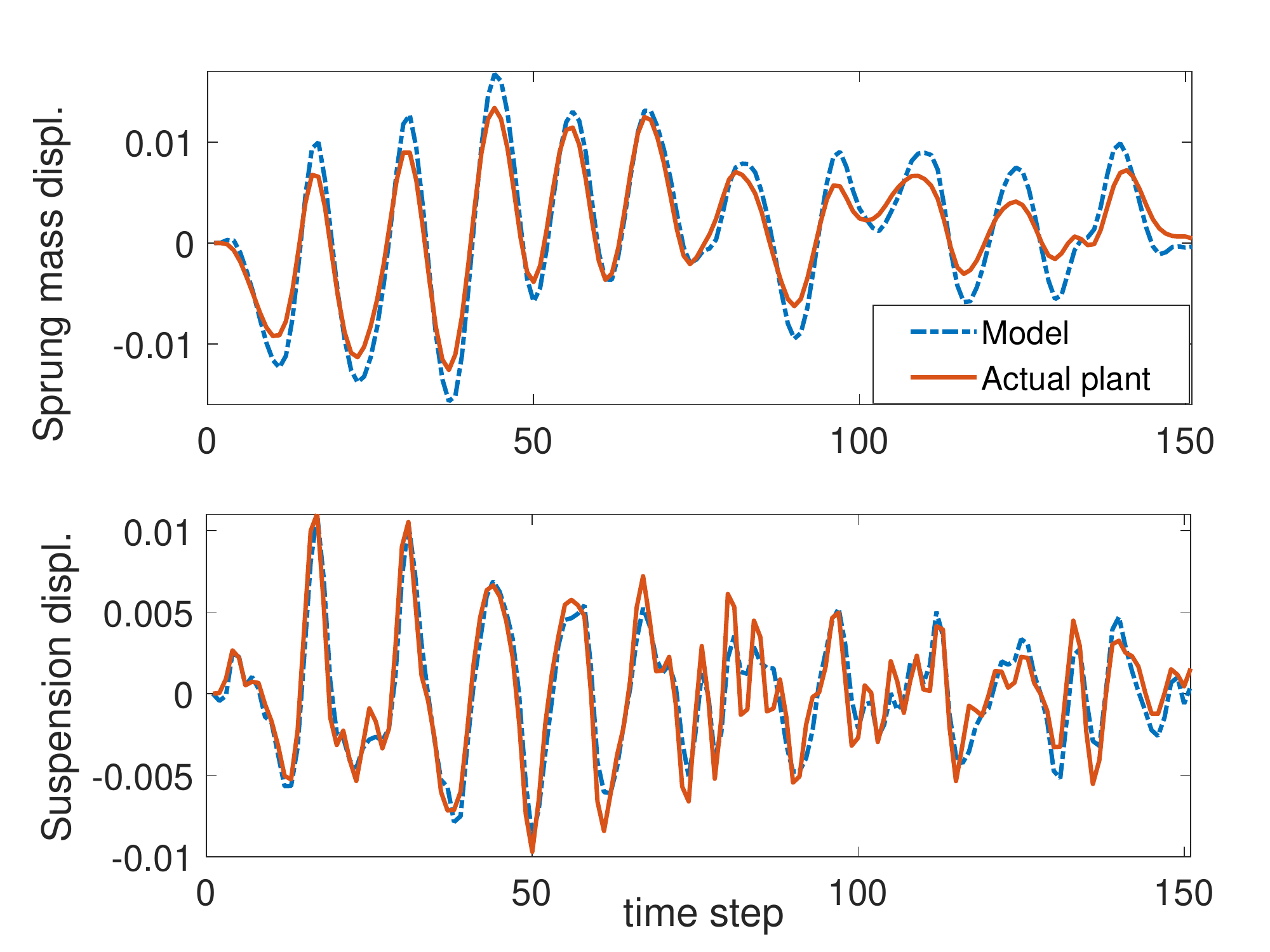}
    \caption{Model evaluation with measurements and predicted output for the first run.}
    \label{fig:y1_y2}
\end{figure}

We first used the model parameters  listed in the Quanser User Manual but the obtained model was fairly inaccurate (maybe due to wear-and-tear).
As such, we fine-tuned the model parameters using the Matlab Parameter Identification toolbox, and the identified parameters for the first experimental setup are included in Table~\ref{paramet_exp}. In addition, the low-pass filter in Eqn.~\ref{eq:filter} with  $a=-5$ and $b=0.0134$ is used to best fit the system setup.
For the experimental test we considered 10 different sets of parameters to simulate 10 heterogeneous vehicles. Since it is difficult to change the damping and stiffness of the suspension station, we used extra masses attached to the sprung mass to change the dynamics of the suspension station. Specifically, the sprung masses representing each vehicle is chosen as
\begin{equation*}
    M_{s,i}=M_{s,1}+(i-1)\times 0.1 \text{ } kg, \text{ } i=1,\dots, 10.
\end{equation*}
For each experiment setup we have re-identified the parameters for different $M_{s,i}$'s which the results are not included here.
Then for each set of sprung mass we performed the cloud-assisted  collaborative estimation as in the simulation section to mimic the collaborative estimation with multiple vehicles. For the first run, the actual measurements from the suspension station and the model output is compared in Fig.~\ref{fig:y1_y2}, which shows decent performance but still some mismatch between the identified model and the actual plant.

For each run, the experiment is carried out for 4.5 seconds with a sampling time of $T_s=0.03$ second. We assume that all vehicles pass the road segment with the same speed. The position uncertainty variance is also chosen as $\sigma_s=0.2$, the same as the one used in the simulation studies. 

\subsubsection{Onboard Estimation Performance}
\begin{figure}[h]
    \centering
    \includegraphics[width=0.8 \linewidth]{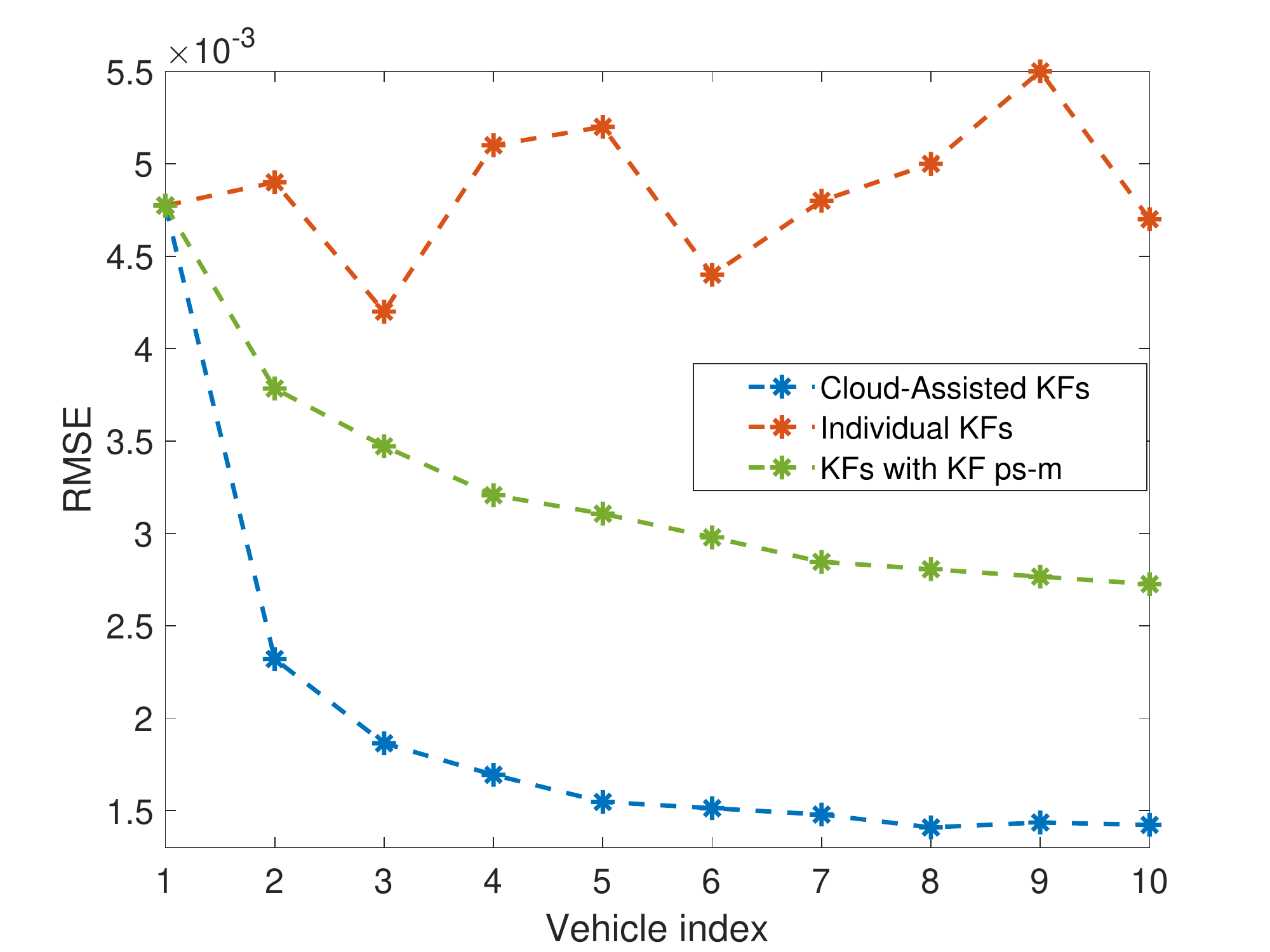}
    \caption{RMSE Comparison of different onboard KF estimation schemes in experiment.}
    \label{fig:Kferrors_exp}
\end{figure}

Fig.~\ref{fig:Kferrors_exp} summarizes the onboard KF estimation performance for the considered experimental setup. As expected,  onboard KFs with GP pseudo-measurements perform better as compared to the case without pseudo-measurements. Also, as seen before, the RMSE decreases after each run, partly due to the improved accuracy of the pseudo-measurements after each NIGP regression. The results also confirm that the effective performance of the framework under heterogeneous vehicles and moderate model uncertainties.

\subsubsection{Cloud NIGP Performance}
\begin{figure}[h]
    \centering
    \includegraphics[width=0.9 \linewidth]{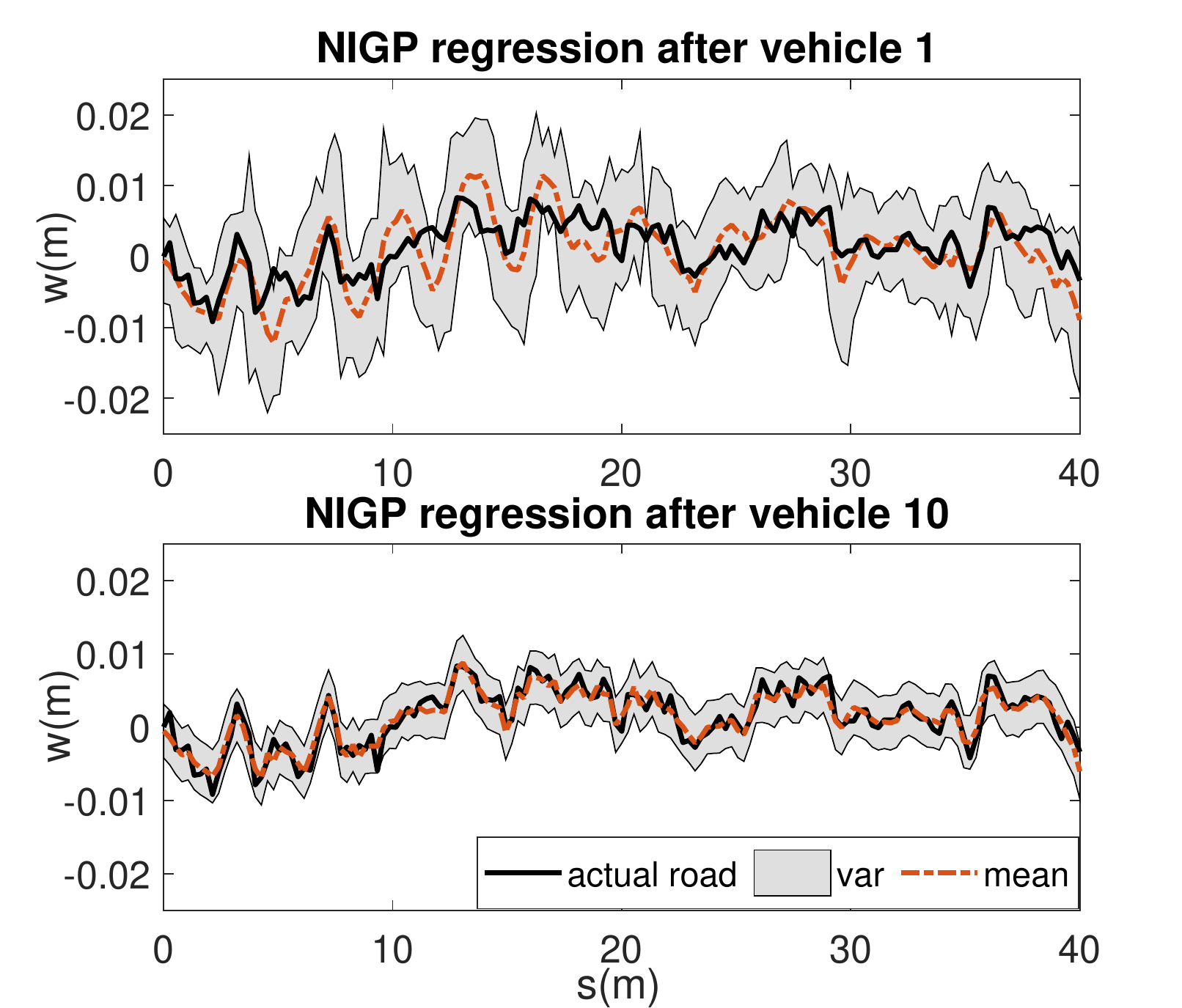}
    \caption{ NIGP regression results  after one run (top) and after 10 runs (bottom). }
    \label{fig:nigpv10_v1}
\end{figure}

\begin{figure}[h]
    \centering
    \includegraphics[width=1 \linewidth]{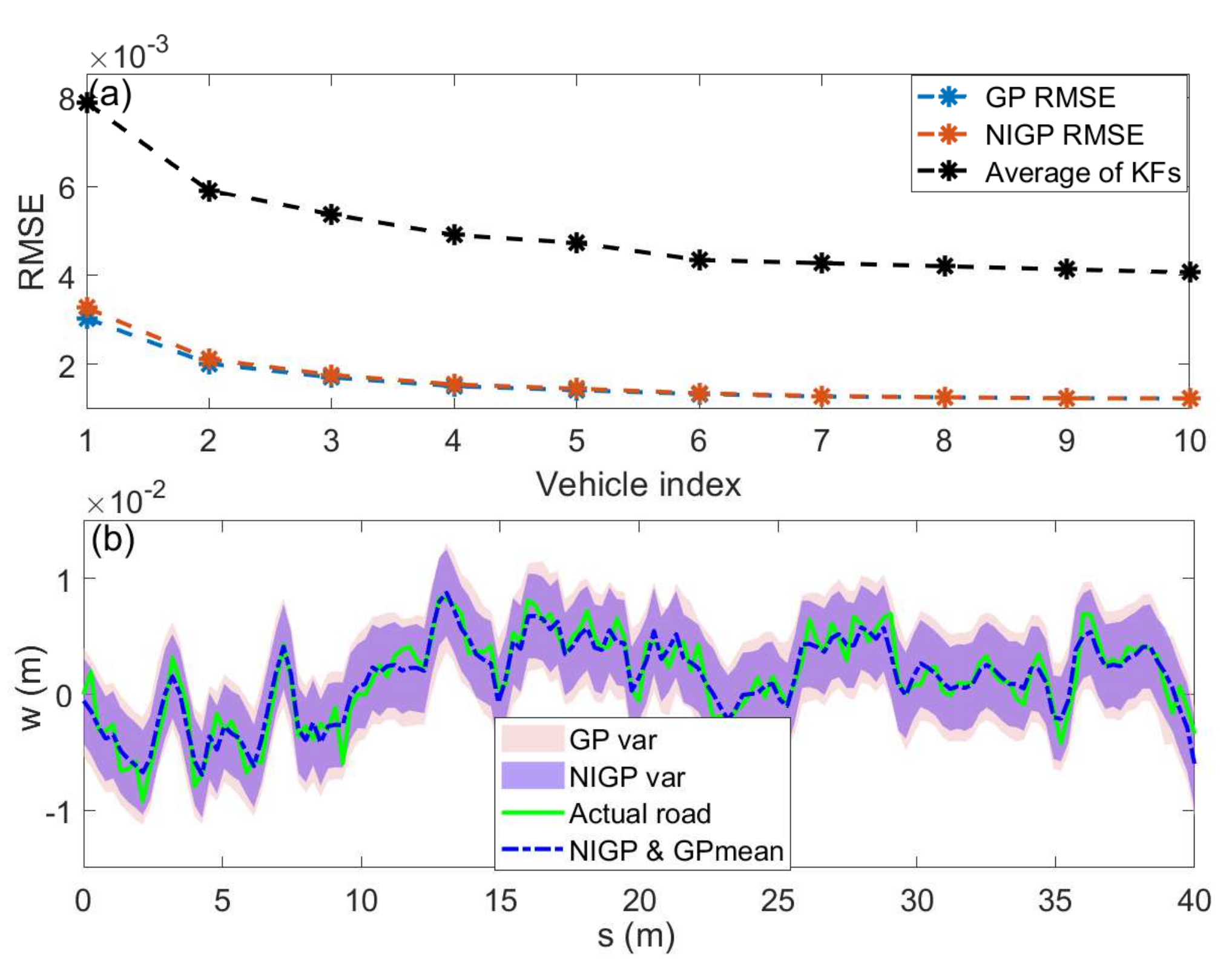}
    \caption{\textcolor{black}{(a): RMSE comparison between NIGP, GP and average of KFs. (b) Comparison of NIGP and GP fit after 10 runs. Only one mean function is shown as the two mean functions are almost identical.}}
    \label{fig:nigp_gp_exp}
\end{figure}

Figs.~\ref{fig:nigpv10_v1}-\ref{fig:nigp_gp_exp} summarize the experimental results on the cloud-based NIGP regression. It can be seen from the top figure in Fig.~\ref{fig:nigpv10_v1} that the first NIGP mean function is not accurate and the variance is large.
With more runs,  the mean function becomes closer to the actual road and the uncertainty reduces, where the final estimate after 10 runs is shown in the bottom figure of Fig.~\ref{fig:nigpv10_v1}. 
Similar to the results from the simulation study, Fig.~\ref{fig:nigp_gp_exp} (a) shows the RMSE comparison among standard GP, NIGP, and a simple average of all KF estimates. It can be seen that the GP variants have similar performance, both significantly outperforming the simple averaging scheme.  However, reduced variance is achieved with NIGP, which is shown in Fig.~\ref{fig:nigp_gp_exp} (b). 

\section{Conclusion}
In this paper, a novel cloud-based collaborative road information discovery framework using multiple heterogeneous vehicles was developed. Gaussian process was used to crowdsource individual estimates, which was then used as pseudo-measurements for future vehicles to enhance its local measurements. We show that this pseudo-measurement strategy was able to greatly enhance the local estimation performance. The enhanced local estimation was then uploaded to the cloud to update the Gaussian process estimation. A noisy-input Gaussian Process method was exploited to explicitly account for GPS uncertainties, which was able to reduce the variance. Extensive simulations and experiments on the application of road profile estimation were  performed to show the efficacy of our framework. Future work will focus on developing more data-efficient Gaussian processes, i.e., without uploading a full set of points for each segment. 

\bibliographystyle{ieeetr}
\bibliography{main}

\begin{IEEEbiography}
 [{\includegraphics[width=1in,height=1.25in,clip,keepaspectratio]{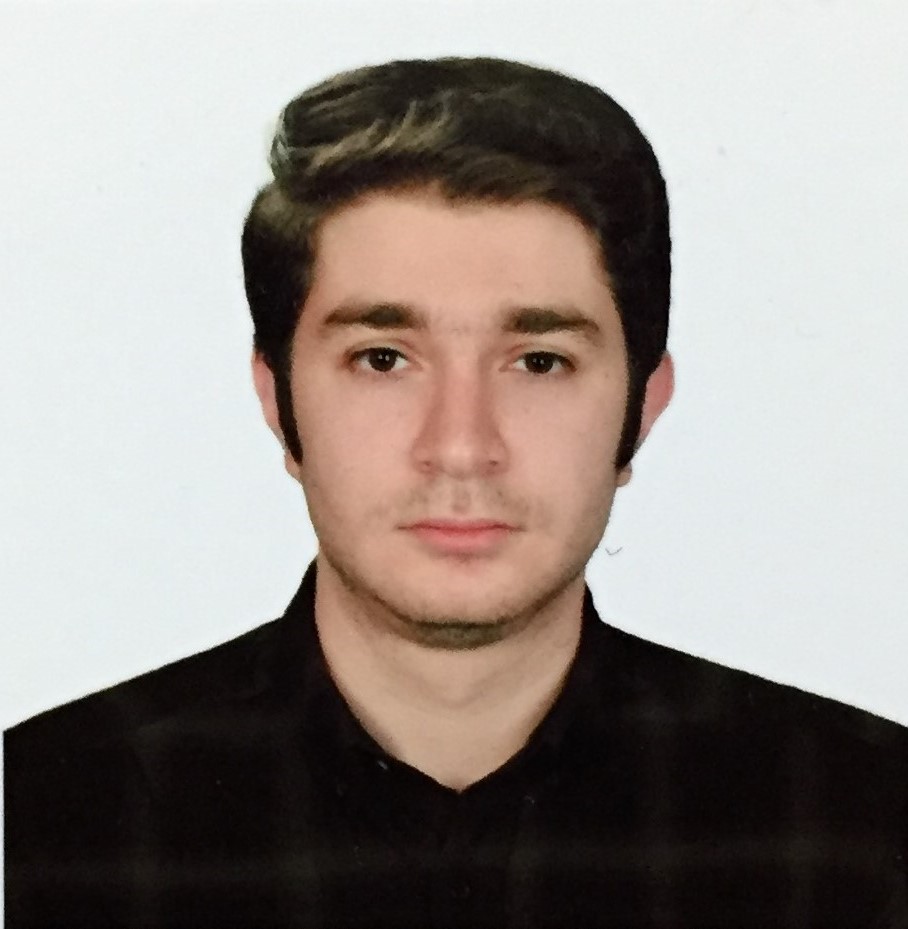}}]{Mohammad R. Hajidavalloo} obtained his B.Sc. and M.Sc. degree from  University of Tehran in Mechanical Engineering  in 2016 and 2018 respectively.

He is currently pursuing the Ph.D. degree in the department of Mechanical Engineering at Michigan State University. His research interests include Learning-Based Control, Model Predictive Control, and Optimal Control, with applications in Connected and Autonomous Vehicles. 
\end{IEEEbiography}

\begin{IEEEbiography}
[{\includegraphics[width=1in,height=1.25in,clip,keepaspectratio]{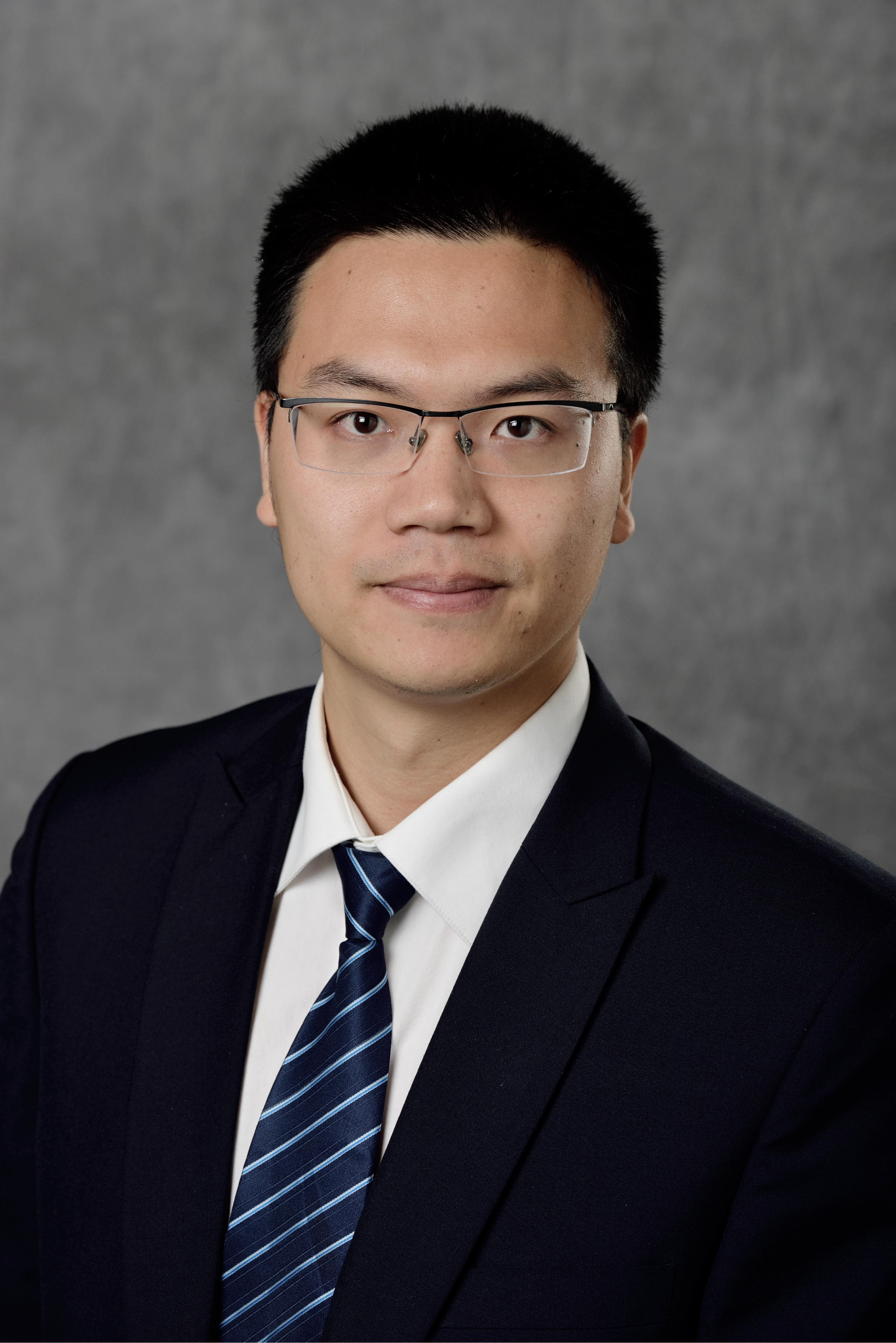}}]{Zhaojian Li}  received his B. Eng. degree from Nanjing University of Aeronautics and Astronautics in 2010. He obtained M.S. (2013) and Ph.D. (2015) in Aerospace Engineering (flight dynamics and control) at the University of Michigan, Ann Arbor.

He is currently an Assistant Professor with the department of Mechanical Engineering at Michigan State University. His research interests include Learning-based Control, Nonlinear and Complex Systems, and Robotics and Automated Vehicles. He is a senior member of IEEE and a recipient of the NSF CAREER Award.
\end{IEEEbiography}

\begin{IEEEbiography}
[{\includegraphics[width=1in,height=1.25in,clip,keepaspectratio]{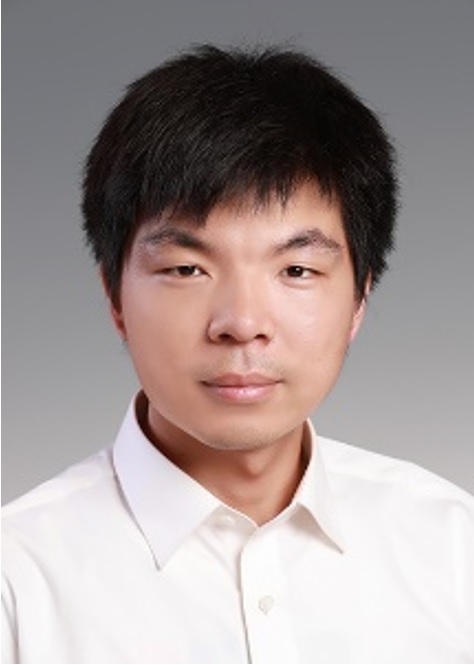}}]{Xin Xia} received the B.E. degree in vehicle engineering from the School of Mechanical and Automotive Studies, South China University of Technology, Guangzhou, China, in 2014, and the Ph.D. degree in vehicle engineering from the School of Automotive Studies, Tongji University, Shanghai, China, in 2019. He was a Postdoctoral Fellow associated with Dr. A. Khajepour with the Department of Mechanical and Mechatronics Engineering, University of Waterloo, Waterloo, ON, Canada, from Jan.2020 to March.2021. He is currently a Postdoctoral researcher associated with Dr.Jiaqi Ma with the Department of Civil and Environmental Engineering, University of California, Los Angeles, CA, USA. His research interest includes state estimation, cooperative localization, cooperative perception, and dynamics control of the autonomous vehicle.
\end{IEEEbiography}

\begin{IEEEbiography}
 [{\includegraphics[width=1in,height=1.25in,clip,keepaspectratio]{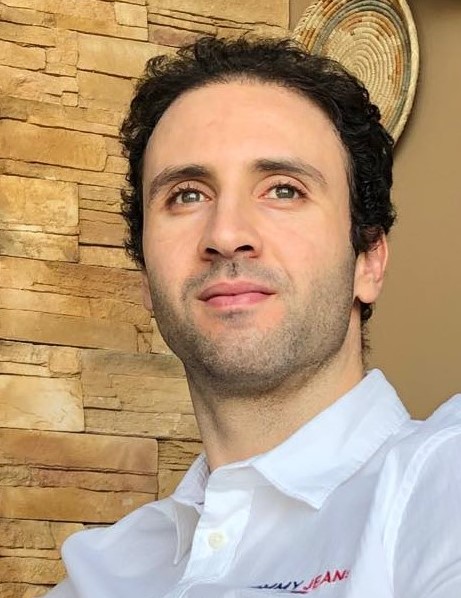}}]{Ali Louati} received the B.Sc. and M.Sc. degrees in
computer science and business from the University
of Sfax, Tunisia, and the Ph.D. degree in computer
science and business from the ISG, University of
Tunis, Tunisia, in 2010, 2013, and 2018, respec-
tively. From 2014 until 2018, he was a research
assistant at both, IFMA, LIMOS, Clermont-Ferrand,
France, and King Saud University. Currently, he is an
assistant professor at Prince Sattam Bin Abdulaziz
University, Saudi Arabia. He is a member of the
SMART Laboratory at ISG, University of Tunis.
He has contributed to several research projects in Tunisia, France, and
Saudi Arabia. His main research interests include machine learning, artificial
immune systems, data mining, and multiobjective optimization. The main
application domain consider by Ali Louati is the development of Intelligent
Transportation Systems. He serves as a reviewer for several international
journals such as IEEE Transactions on Intelligent Transportation Systems, Automatica, and Applied Soft Computing.
\end{IEEEbiography}

\begin{IEEEbiography}
 [{\includegraphics[width=1in,height=1.25in,clip,keepaspectratio]{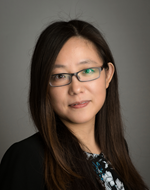}}]{Minghui Zheng} received the B.E. and M.E. degrees, in 2008 and 2011 respectively, from Beihang University, Beijing, China, and the Ph.D. degree in Mechanical Engineering, in 2017, from University of California, Berkeley, USA.  She joined University at Buffalo, NY, USA, in 2017, where she is currently an assistant professor in Mechanical and Aerospace Engineering. Her research interests include advanced learning, estimation, and control with applications to high-precision and robotic systems.
\end{IEEEbiography}


\begin{IEEEbiography}
 [{\includegraphics[width=1in,height=1.25in,clip,keepaspectratio]{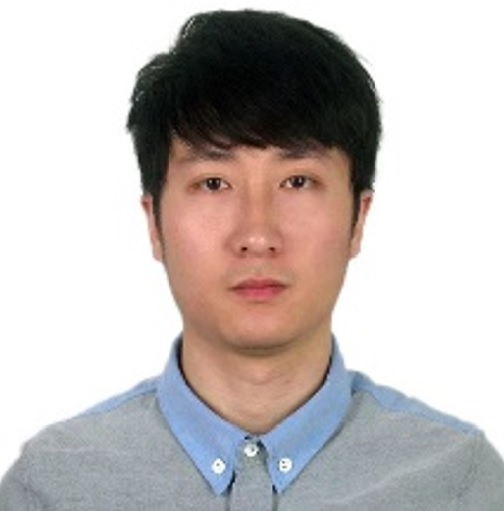}}]{Weichao Zhuang}  received the B.Eng. and Ph.D. degrees in mechanical engineering from the Nanjing University of Science and Technology, Nanjing, China, in 2012 and 2017, respectively. From January 2014 to December 2015, he was a Visiting Student with the Department of Mechanical Engineering, University of Michigan, Ann Arbor, MI, USA. He is currently an Associate Professor with the School of Mechanical Engineering, Southeast University, Nanjing, China. His current research interests include connected and automated vehicles, optimal control, and multi-agent control.
\end{IEEEbiography}

\end{document}